\documentclass[11pt]{article}
\usepackage[top=2cm,left=2cm,right=2cm]{geometry}
\usepackage[numbers,square]{natbib}

\usepackage{float}
\usepackage[english]{babel}
\usepackage{epsfig}

\usepackage{hyperref}
\hypersetup{
    colorlinks,
    citecolor=green,
    filecolor=black,
    linkcolor=blue,
    urlcolor=black
}
\usepackage{amssymb}
\usepackage{amsfonts}
\usepackage{epsf}
\usepackage{rotating}
\usepackage{graphicx}
\usepackage{amsmath}
\usepackage{fancyhdr}
\usepackage{subfigure}
\usepackage{babel}
\usepackage{graphics}
\usepackage{geometry}
\usepackage{pstricks}
\usepackage{color}

\makeatletter
\def\cleardoublepage{\clearpage\if@twoside
\ifodd\c@page
\else\hbox{}\thispagestyle{empty}\newpage
\if@twocolumn\hbox{}\newpage\fi\fi\fi}
\makeatother

\let\a=\alpha   \let\b=\beta   \let\g=\gamma   
         \let\q=\theta
        
           \let\p=\pi      
\let\s=\sigma        

\def\sq{\,\raise.5pt\hbox{$\nbox{.09}{.09}$}}
\def\sqb{\,\raise.5pt\hbox{$\overline{\nbox{.09}{.09}}$}}

\def\a{\alpha}
\def\b{\beta}

\def\g{\gamma}

\def\ds#1{#1\kern-1ex\hbox{/}}
\def\dsh{h\kern-1.2ex /}

\newcommand{\bea}{\begin{eqnarray}}
\newcommand{\eea}{\end{eqnarray}}

\def\nn{\nonumber}

\def\beq{\begin{equation}}
\def\eeq{\end{equation}}

\def\ba{\begin{eqnarray}}
\def\ea{\end{eqnarray}}

\newcommand{\beqa}{\begin{eqnarray}}
\newcommand{\eeqa}{\end{eqnarray}}

\newcommand{\pd}{\partial}

\newcommand{\bes}{\begin{subequations}}
\newcommand{\ees}{\end{subequations}}

\begin{document}

\begin{center}

{\bf \large Electroweak Corrections to Photon Scattering, Polarization and Lensing \\ in a Gravitational Background 
and the Near Horizon Limit}

\vspace{0.2cm}
\vspace{0.3cm}

\vspace{1.5cm}
{\bf $^{(1)}$Claudio Corian\`{o}, $^{(1)}$Luigi Delle Rose, $^{(1)}$Matteo Maria Maglio and $^{(1,2)}$Mirko Serino }\footnote{claudio.coriano@le.infn.it, 
luigi.dellerose@le.infn.it, matteomaria.maglio@le.infn.it, mirko.serino@ifj.edu.pl}

\vspace{1cm}

{\it $^{(1)}$Dipartimento di Matematica e Fisica "Ennio De Giorgi", 
Universit\`{a} del Salento and \\ INFN-Lecce, Via Arnesano, 73100 Lecce, Italy}\\
\vspace{0.5cm}
{\it $^{(2)}$Institute of Nuclear Physics\\
Polish Academy of Sciences, ul. Radzikowskiego 152
31-342 Krakow, Poland}\\

\vspace{.5cm}
\begin{abstract} 
We investigate the semiclassical approach to the lensing of photons in a spherically symmetric gravitational background, starting from Born level and include in our analysis the radiative corrections obtained from the electroweak theory for the graviton/photon/photon vertex.  In this approach, the cross section is related to the angular variation of the impact parameter ($b$), which is then solved for $b$ as a function of the angle of deflection, and measured in horizon units ($b_h\equiv b/(2 G M)$). Exact numerical solutions for the angular deflection are presented.  The numerical analysis shows that perturbation theory in a  weak background agrees with the classical Einstein formula for the deflection already at distances of the order of $20$ horizon units ($\sim 20\, b_h$) and it is optimal in the description both of very strong and weak lensings.  We show that the electroweak corrections to the cross section are sizeable, becoming very significant for high energy gamma rays. Our analysis covers in energy most of the photon spectrum, from the cosmic microwave background up to very high energy gamma rays, and scatterings with any value of the photon impact parameter. We also study the helicity-flip photon amplitude, which is of $O(\alpha^2)$ in the weak coupling $\alpha$, and its massless fermion limit, which involves the exchange of a conformal anomaly pole. The corresponding cross section is proportional to the Born level result and brings to a simple renormalization of Einsten's formula.
\end{abstract}
\end{center}

\newpage
\section{Introduction}
Photon scattering \cite{Berends:1975ah} and lensing by gravity has been investigated along the years with an ever increasing interest both in astrophysics and cosmology. Since Einstein's original work, this topic has developed into a very important area of research, providing both a method to test Einstein's general 
relativity (GR), and also a way to probe the distribution of matter and dark matter in the universe \cite{Massey:2010hh}. 
Studies of weak lensing \cite{Wambs:1998, Schneider:2005ka}, for instance, which are relevant in the latter case, concern the identification of small deflections of the photon paths due to gravity. In this case they are of the order of 1 arcsecond or so, and are investigated on large statistical galaxy samples in order to provide information on the underlying distribution of matter or dark matter on very large scales.
On the other hand, larger angles of deflection ($\sim 20$ arcsecs), classified as deflections due to strong lensing, are expected to occur when a photon trajectory nears a large concentration of matter. Obviously, even stronger lensing effects are felt by the photons when these approach the horizon of a black hole (very strong lensing), where the angular deflection of an incoming beam may be of few degrees or even larger.  
In the limiting case of a beam grazing the photon sphere of a black hole, the photons may perform one or more turns around the deflector before reaching the asymptotic region, a phenomenon which renders these studies more involved. In this case, in fact, the lens provides two infinite sets of images (see the discussion in \cite{ Bozza:2001xd}). Analytical and numerical studies of black hole lenses have covered Schwarzschild, Reissner-Nordstr\"om and other metric solutions, such as spherically symmetric and rotating geometries, braneworld geometries, naked singularities etc. (see for instance \cite{Virbhadra:1999nm,Virbhadra:2002ju, Bozza:2001xd, BinNun:2009jr} and refs. therein).
 
While the classical deflection of a photon path by gravity is a well-studied 
aspect of GR, the explicit form of the full electroweak corrections have never been discussed before. In particular, they have been limited only to the case of QED and for weak lensing \cite{Berends:1975ah}. With no doubt these corrections are small, but grow in size as one 
approaches scattering centers of larger gravitational force and at high energy. In general, a gravitational cross section grows quadratically with the mass $M$ of the source, here assumed to be of the order of the solar mass $M_\odot$ or larger. At the same time, the $\log(E)$ and $\log(E)^2$ dependence in the perturbative expansion is responsible for the growth with the energy $E$ of the corresponding K-factors at 1-loop, being the leading order cross section independent of energy. Such is the case of high energy/very high energy gamma rays, which are part of the photon spectrum at cosmological level, covering energies up to $10^{18}$ eV, a region known as "the ankle" of the cosmic ray spectrum. These may originate from primary protons of very high energy, via the mechanisms of pion photoproduction, or by the inelastic nuclear collisions of these primaries, close to their original sources. The possible extragalactic origin of these very energetic cosmic rays, especially for energies above the TeV region, which are expected to be accompanied by almost equally energetic gamma emissions, is nowadays widely debated \cite{Aab:2014bha, Apel:2013ura, Ahlers:2013xia}. 

In the case of photons, one important point which, we believe, has not been sufficiently investigated up to now, is the relation between the quantum and the classical predictions of their scattering and lensing. For instance, the study of their deflection has mostly been limited to the case of classical GR and given by Einstein's expression. We recall that Einstein's formula for the deflection of light (see \cite{Amore:2006pi, Amore:2006xp}) finds application in the context of gravitational lensing - both strong and weak - being at the basis of the gravitational lens equation. However, the comparison between the two descriptions, first at Born level and then with the inclusion of the corresponding radiative corrections, is of extraordinary interest. It involves a semiclassical limit which needs to be investigated with care. The study of this limit and the determination of the region where classical and semiclassical approaches, mediated by the quantum corrections, share similar predictions, motivates our analysis.

A way to compare classical and quantum predictions was suggested long ago by Delbourgo and Phocas-Cosmetatos \cite{Delbourgo:1973xe}. In their work, the authors equate the quantum mechanical cross section, computed in ordinary perturbation theory,
with the classical one, expressed in terms of the impact parameter of the incoming photon, treated as a classical particle 
\beq
\frac{b}{\sin\theta} \, \vline  \frac{d b}{d\theta}\vline = \frac{d \sigma}{d\Omega}.
\label{semic}
\eeq
The expression above defines a differential equation for the impact parameter whose solution relates $b$ to $\theta$, the classical angle of deflection. This allows a comparison between the two approaches, giving a deflection which is in agreement with Einsten's prediction in the case of weak lensing.\\ 
 As we have already mentioned, in the presence of radiative corrections, the angle of deflection becomes energy dependent, generating a true gravitational rainbow \cite{Delbourgo:1973xe, Golowich:1990gp, Accioly:2004bm}. This feature, which is absent both in the classical case and in the quantum case at Born level, is an important effect which sets a distinction between the classical and the quantum approaches to lensing. Obviously, its phenomenological impact is quite small in size, unless one can show that the semiclassical description can be extended also to regions which are far closer to the event horizon of a compact massive source. \\
In \cite{Delbourgo:1973xe} the authors also derived an effective vertex describing the local operator emerging from the exchange of a charged boson or fermion in the loop of spin $J$, given by the interaction 
\beq
-\frac{1}{4} g^{\mu\nu}\left[ g^{\kappa\lambda}F_{\mu\kappa}F_{\nu\lambda} + \frac{\alpha}{720 m^2} (-1)^{2 J}(2 J +1)F_{\kappa\lambda} 
\stackrel{\leftrightarrow}{\partial_\mu}\stackrel{\leftrightarrow}\partial_\nu F_{\kappa\lambda}\right],
\label{delb}
\eeq
where $m$ the mass of the virtual particle. Eq.(\ref{delb}) was then used in Eq.(\ref{semic}) to predict the photon deflection in the case of weak lensing. \\
It should be remarked that the 
effective vertex defined above is only valid at the lowest order in an expansion in $t/m^2$, where $t$ is the momentum transfer of the graviton and $m$ is the mass of the lightest virtual particle exchanged in the loops. As such, it does not account for the full one-loop corrections and cannot be used for a wider study of the deflection. In any case, Eq.(\ref{delb}) is expected to provide a realistic description of the photon deflection only in the limit of weak deflection, where the typical momentum transfers of the graviton cover the far infrared region. Such are those scatterings characterized by very large impact parameters, as we are going to specify in detail below. 
Obviously, the approach becomes inaccurate once we get very close to the Schwarzschild radius ($R_S=2\, G M$) of a compact source, while the full perturbative solution of Eq.(\ref{semic}) is expected to be predictive at distances of the order of $10^6\, R_S$, which are comparable with the radius of the sun. \\ 
Our study shows that the range of validity of perturbation theory in the presence of a weak gravitational background is not only limited to very large distances, very far from the event horizon, as known before, but becomes realistic also at distances which are much closer to it. It is essential, for this result to hold, to include in the numerical analysis the entire dependence on the momentum transfer $t$ of the graviton/photon/photon vertex. \\ 
In the case of infrared photons, for instance for photons of the CMB, the typical momentum transfers involved in the interaction are quite small, at least for scattering far from the horizon, and one is allowed to perform a suitable expansion in $t/m^2$, with $m$ being, in the SM, the electron mass. A first order expansion in this parameter is sufficient to obtain very accurate results both for the differential cross section 
and for the semiclassical deflection obtained via Eq.(\ref{semic}).
On the other hand, closer to the horizon of the source, photons of the CMB are expected to deflect significantly, and the inclusion of the additional contributions in $t/m^2$, which are generated by the radiative corrections, is therefore mandatory.  As we are going to show, 
at the other end of the spectrum - for very high energy (VHE) photons - the numerical analysis of Eq.(\ref{semic}) is quite straightforward to perform, and the convergence of the differential equation for the impact parameter at any value of $b$ is rather optimal.

\subsection{Content of our work}
Our work is organized as follows. In the next few sections we are going to briefly overview the formalism of photon scattering in a weak background, defining our conventions. The interaction is worked out for the case of an external body of heavy mass ($M$), treated as a static point-like source, which is the case relevant for photon-gravity interactions off the external core of the source. Some modifications of this approach which take into account the geometric form factor of the source are necessary if a particle is allowed to interact with the interior region, such as for a neutrino or a dark matter fermion, 
but they are not relevant in the photon case. By using 
the impact parameter cross section and its solution $b$ as a function of the angle of deflection, we discuss the inversion of Eq.(\ref{semic}), proceeding with several numerical studies.
We then present the radiative corrections to the amplitude and to the cross section. We then turn to extract the expression of $b$ to one-loop order in the complete electroweak theory, determining the structure of the angular deflection at this order. \\
The background metric, over which we expand, is the retarded solution of the linearized Einstein's equations and coincides with the Schwarzschild metric, once we take its expression in the limit of a weak external field. \\ We will investigate both the polarized and the unpolarized cross sections, and present results for these over a very wide range of energy. The photon spectrum, which we are interested in, will cover in our analysis both the case of infrared photons and of ultra high energy gamma's. We will give particular attention to the helicity-flip cross section, which is quite small compared to the helicity-preserving component, but grows significantly as we increase the energy ($\sim E^4$), at least for small values of $E$, finally reaching a plateau at high energy. As for any gravitational cross section, it grows quadratically with the mass of the source, becoming relevant for scatterings off massive/ supermassive black holes. This component of the unpolarized cross section appears at $O(\alpha^2)$ and carries information on the anomaly form factor of the graviton/photon/photon vertex.     

\section{Definitions and conventions} %

We will be following the same conventions already used in a previous investigation \cite{Coriano:2011zk}, where the action describing the dynamics of the SM in a gravitational background is defined as
\beq
\mathcal{S} =  \mathcal{S}_{SM} +\mathcal{S}_G + \mathcal{S}_{I}
\eeq
with $\mathcal{S}_{SM}$ denoting the SM Lagrangian, while 
\bea
\mathcal{S}_G  
&=& 
- \frac{1}{\kappa^2}\int d^4 x \sqrt{-{g}}\, R \nn\\
\mathcal{S}_I  
&=& 
\chi \int d^4 x \sqrt{-{g}}\, R \, H^\dag H 
\label{thelagrangian}
\eea
denote respectively the gravitational Einstein term, with $\kappa^2 = 16\,\pi\,G$
and a term of improvement $\mathcal{S}_I$ involving the Higgs doublet $H$.
The latter, for the conformal value $\chi=1/6$, is responsible for generating a symmetric and traceless 
energy-momentum tensor (EMT) and guarantees the renormalizability of Green functions containing external gravitons and an arbitrary set of SM fields.
We will be using the flat metric $\eta_{\mu\nu}= \textrm{diag}(1,-1,-1,-1)$, with an expansion of the form
\bea
 g_{\mu\nu}=\eta_{\mu\nu} + \kappa \, h_{\mu\nu} + O(\kappa^2) \,.
\eea
The energy momentum tensor (EMT) 
is computed for the SM by embedding the corresponding Lagrangian to a curved spacetime by the relation
\beq
T_{\mu\nu}=\frac{2}{\sqrt{-g}}\frac{\delta \left(S_{SM}+S_{I}\right)}{\delta g^{\mu\nu}} \bigg|_{g=\eta} \,.
\eeq 
We recall, at this point, that the fermions are coupled to gravity using the spin connection $\Omega$ induced 
by the curved metric $g_{\mu\nu}$. This allows to define a spinor derivative $\mathcal{D}$ which transforms covariantly under local 
Lorentz transformations. If we denote with $\underline{a},\underline{b}$ the Lorentz indices of a local free-falling frame, and with
$\sigma^{\underline{a}\underline{b}}$ the generators of the Lorentz group in the spinorial representation, the spin connection takes 
the form
\beq
 \Omega_\mu(x) = 
\frac{1}{2}\sigma^{\underline{a}\underline{b}}V_{\underline{a}}^{\,\nu}(x)V_{\underline{b}\nu;\mu}(x)\, ,
\eeq
where we have introduced the vielbein $V_{\underline{a}}^\mu(x)$. The covariant derivative of a spinor in a given representation
$(R)$ of the gauge symmetry group, expressed in curved $(\mathcal{D}_{\mu})$ coordinates is then given by
\beq \mathcal{D}_{\mu} = \frac{\pd}{\pd x^\mu} + \Omega_\mu   + A_\mu,\eeq
where $A_\mu\equiv A_\mu^a\, T^{a}_{(R)}$ are the gauge fields and $T^{a}_{(R)}$ the group generators,
giving a Lagrangian of the form
\bea 
\mathcal{L}
&=& 
\sqrt{-g} \bigg\{\frac{i}{2}\bigg[\bar\psi\g^\mu(\mathcal{D}_\mu\psi)
 - (\mathcal{D}_\mu\bar\psi)\g^\mu\psi \bigg] - m\bar\psi\psi\bigg\}.       
\eea
The full EMT is given by a minimal tensor $T^{Min}_{\mu\nu}$ (without improvement) and a term of improvement, $T^I_{\mu\nu}$, 
generated by the conformal coupling of the scalars
\bea
T_{\mu\nu} = T^{Min}_{\mu\nu} + T^I_{\mu\nu} \,,
\eea
where the minimal tensor is decomposed into 
\bea
T^{Min}_{\mu\nu} =
T^{f.s.}_{\mu\nu} + T^{ferm.}_{\mu\nu} + T^{Higgs}_{\mu\nu} + 
T^{Yukawa}_{\mu\nu} + T^{g.fix.}_{\mu\nu} + T^{ghost}_{\mu\nu} \, 
\eea
in terms of the field strength contributions (f.s.), the Higgs and Yukawa terms, the gauge-fixing terms (g.fix.) and the 
ghost contributions, whose expressions have been given in \cite{Coriano:2011zk}. Notice that $T^I_{\mu\nu}$ is defined modulo the constant $(\chi)$ of $\mathcal{S}_I$, as defined in Eq.(\ref{thelagrangian}). \\
The evaluation of all the vertices describing the coupling of this operator to all the fields of the SM is rather cumbersome. 
For this  reason, in a previous work we have derived a large set of Ward and Slavnov-Taylor identities (STI's) for correlators involving a $T_{\mu\nu}$ insertion, in order to secure the consistency of the explicit expressions of all the one-loop corrections. The STI's have been 
obtained in the $R_\xi$ gauge, which is our chosen gauge. Details of this analysis in the neutral current sector of the SM can be found in \cite{Coriano:2011zk}. The analyses in the QED and QCD cases have been performed, respectively, in \cite{Armillis:2009pq} and \cite{Armillis:2010qk}. \\
As a side remark on our notations, we will be denoting with $\theta_d$ the angular deflection as predicted by the classical geodetic 
equation of GR.  The solution of (\ref{semic}), which relates the impact parameter to the semiclassical angle of deflection, will be indeed denoted as 
$b=b(\theta_d)$. Instead, we will denote with $\theta$ the scattering angle of the quantum differential cross section. The transition from $\theta$ to $\theta_d$ requires the integration of (\ref{semic}), with $\theta_d$ being the boundary of the angular integration in $\theta$ of $d\sigma/d\Omega$, as discussed in Section 7.

\section{The external source}
In this section we proceed with the analysis of the fluctuations of the metric in the Schwarzschild case. 
We will consider the potential scattering of a photon off an external static source, which acts as a
perturbation on the otherwise flat spacetime background. The source is characterized by an EMT $T^{ext}_{\mu\nu}$ and the fluctuations are determined by solving the linearized equations of GR. 
These take the form
\beq
\square\left(h_{\mu\nu}-\frac{1}{2}\eta_{\mu\nu} h\right)=-{\kappa}T^{ext}_{\mu\nu} \,,
\eeq
where $h\equiv h^{\mu\nu}\eta_{\mu\nu}$, and can be rewritten as
\beq
\square h_{\mu\nu}={\kappa}\, S_{\mu\nu}, \qquad \mbox{with} \qquad S_{\mu\nu}=
- \left( T^{ext}_{\mu\nu}-\frac{1}{2}\eta_{\mu\nu}T^{ext}\right) \,.
\label{oneq}
\eeq
The external field $h_{\mu\nu}$ is obtained by convoluting the static source with the retarded propagator 
\beq
G_R(x,y)= \frac{1}{4 \pi}\, \frac{\delta(x_0-y_0-|\vec{x}-\vec{y}|)}{|\vec{x}-\vec{y}|},
\eeq
normalized as
\beq
\square G_R(x,y)=\delta^4(x-y). 
\eeq
The solution of Eq.(\ref{oneq}) takes the form 
\beq
h^{ext}_{\mu\nu }(x)={\kappa}\, \int d^4 y\, G_R(x,y)\, S_{\mu\nu}(y)
\eeq
with the EMT of the external localized source, defining $S_{\mu\nu}$, given by
\beq
T^{ext}_{\mu\nu}= \frac{P_{\mu}P_{\nu}}{P_0}\delta^3(\vec{x}) \,.
\eeq
For a compact source of mass $M$ at rest at the origin, with $P_{\mu}=(M,\vec{0}$), we have
\beq
T^{ext}_{\mu\nu}=M\delta^0_\mu\delta^0_\nu\delta^3(\vec{x})
\eeq
which gives
\beq
S_{\mu\nu}=\frac{M}{2}\bar{S}_{\mu\nu} \qquad 
\bar{S}_{\mu\nu}\equiv \eta_{\mu\nu}-2 \delta^0_{\mu}\delta^0_{\nu}
\eeq
and 
\bea
h^L_{\mu\nu}(x)
&=& \frac{2 G M}{\kappa |\vec{x}|}\bar{S}_{\mu\nu},
\label{hh}
\eea
where the field generated by a local (point-like, $L$) mass distribution has the typical $1/r$  $(r\equiv |\vec{x}|)$ behaviour.
The fluctuations are normalized in such a way that $h_{\mu\nu}$ has mass dimension 1, as an ordinary bosonic field, with 
$\kappa$ of mass dimension ${-1}$. Using the expression above, the full metric takes the form  
\beq
ds^2\approx\left(1- \frac{2 G M}{|\vec{x}|}\right)dt^2 -\left(1 + \frac{2 G M}{|\vec{x}|}\right)d\vec{x}\cdot d\vec{x}.
\label{SCH3}
\eeq
This metric coincides with that of Schwarzschild, once we perform on the latter the weak field limit. For this purpose 
we recall that, in the case of a spherically symmetric and stationary source, the ansatz for the metric in polar coordinates  takes the form 
\beq
\label{SCH1}
ds^2=e^{\nu(r)} d t^2 - e^{\lambda(r)} dr^2 - r^2 d\theta - r^2\sin^2\theta d\phi^2 
\eeq
in terms of two undeterminate functions which depend on a radial coordinate $r$, $\nu(r)$ and $\lambda(r)$.
The Einstein equations in the vacuum determine these two functions in terms of a single parameter $C$ in the form 
\beq
e^{\lambda(r)}=\frac{1}{1 -\frac{C}{r}}, \qquad e^{\nu(r)}=1-\frac{C}{r}
\label{SCH2}
\eeq
as
\beq
\label{ds0}
ds^2=(1-\frac{C}{r}) dt^2 - \frac{1}{1 -\frac{C}{r}} dr^2 - r^2 d\theta - r^2\sin^2\theta d\phi^2 \,.
\eeq
Using the change of variable in the radial coordinate
\beq
r=r'\left(1 + \frac{C}{4 r'}\right)^2 ,
\eeq
it is possible to rewrite the vacuum solution in Eq.(\ref{SCH1}) in the form 
\beq
ds^2=\left( \frac{1- \frac{C}{4 r'}}{1 + \frac{C}{4 r'}}\right)^2 dt^2 - \left( 
1 + \frac{C}{4 r'}\right)^4 (dr'^2 + r'^2 d\theta^2 + r'^2 \sin\theta^2 d\phi^2),
\label{SCH4}
\eeq
which is characterized by a single factor multiplying the spatial part $(d\vec{x}\cdot d\vec{x})$, as in the retarded solution given in Eq.(\ref{SCH3}). Taking the limit $r'\to \infty$, which corresponds to the weak field approximation 
(i.e $b_N\equiv GM/(R c^2) \ll 1$), Eq.(\ref{SCH4}) becomes 

\beq
ds^2\approx\left(1- \frac{C}{r'}\right) -\left(1 + \frac{C}{r'}\right) ( dr'^2 + r'^2 d\theta^2 + r'^2 \sin\theta^2 d\phi^2)
\label{SCH5},
\eeq
which allows us to identify $C=2 G M$ and coincides with Eq.(\ref{SCH3}). \\
At this point, moving to momentum space,
the Fourier transform of $h_{\mu\nu}$ in Eq.(\ref{hh}) is defined by 
\bea
 h_{\mu\nu}(q_0,\vec{q})&=& \int d^4 x\, e^{i q\cdot x}\, h_{\mu\nu}(x) \,,
\eea
which for a static field reduces to
 \beq
 h_{\mu\nu}(q_0,\vec{q})= 2\, \pi\, \delta(q_0) h_{\mu\nu}(\vec{q} ),
 \eeq
 with
 \beq
 h_{\mu\nu}(\vec{q})\equiv h_0(\vec{q}) \bar{S}_{\mu\nu}.
 \eeq
The expression above shows that the metric fluctuation in momentum space is entirely defined in terms of the scalar form factor $h(\vec{q})$ which allows to rewrite it in the form
\beq
h_{\mu\nu}(q_0,\vec{q})= 
2 \pi \delta(q_0) \frac{2 G M}{\kappa} \bar{S}_{\mu\nu}\int d^3 \vec{x} 
\frac{e^{i\vec{q}\cdot \vec{x}}}{|\vec{x}|}
= 
2\,\pi\, \delta(q_0) \left(\frac{\kappa M}{2\vec{q}^2}\right)\bar{S}_{\mu\nu}
\label{trans}
\eeq
with  
\beq
 h_0(\vec{q})\equiv \left(\frac{\kappa M}{2 \vec{q}^2}\right), \qquad  h_{\mu\nu}(\vec{q})
\equiv \left(\frac{\kappa M}{2 \vec{q}^2}\right) \bar{S}_{\mu\nu}.
 \label{h0}
 \eeq

\section{The leading order cross section}\label{TreeLevel}
We proceed with the leading order analysis of the photon/graviton interaction, using the tree-level graviton/photon/photon vertex.
For this purpose we denote with $\hat{T}^{(0)\, \mu\nu}$ the tree-level matrix element characterizing the transition amplitude between the initial and the final state photon, in the presence of the gravitational background, mediated by the insertion of the EMT. 
This is given by 
\beq
- i\, \frac{\kappa}{2}\, \hat{T}^{(0)\,\mu\nu} =
V^{\mu\nu\alpha\beta}(p_1,p_2)A^i_\alpha(p_1)\, A^f_\beta(p_2) \, ,
\eeq
where $V^{\mu\nu\alpha\beta}(p_1,p_2)$ denotes the on-shell graviton/photon/photon vertex in momentum space,
having labeled the momenta of the incoming and outgoing photons with $p_1$ and $p_2$ respectively.
It is explicitly given by
\bea
&&
V^{\mu\nu\alpha\beta}(p_1,p_2) =
- i \frac{\kappa}{2} \bigg\{ - p_1 \cdot p_2  \,  C^{\mu\nu\alpha\beta} + D^{\mu\nu\alpha\beta}(p_1,p_2)\bigg\}
\nn \\
&& 
C^{\mu\nu\alpha\beta} = 
\eta^{\mu\alpha}\, \eta^{\nu\beta} + \eta^{\mu\beta} \, \eta^{\nu\alpha}- \eta^{\mu\nu} \, \eta^{\alpha\beta}\, , \nn
\\
&& 
D^{\mu\nu\alpha\beta} (p_1, p_2) =
- \eta^{\mu\nu} \, p_1^{ \beta}\, p_2^{\alpha} + \biggl[ \eta^{\mu\beta} p_1^{\nu} p_2^{\alpha}
+ \eta^{\mu\alpha} \, p_1^{\beta} \, p_2^{\nu} - \eta^{\alpha\beta} \, p_1^{\mu} \, p_2^{\nu}
+ (\mu \leftrightarrow \nu)\biggr] .
\eea
The incoming and outgoing plane waves have been defined as 
\beq
A^i_{\alpha}(p_1) = \mathcal{N}_i  \, \epsilon_{\alpha}(p_1)\, ,
\qquad
A^f_{\alpha}(p_2) = \mathcal{N}_f  \, \epsilon^*_{\alpha}(p_2)\, ,
\qquad
\mathcal{N}_i = \sqrt{\frac{1}{2\, E_1\, V }}\, , \qquad \mathcal{N}_f = \sqrt{\frac{1}{2\, E_2\, V }}\,,
\label{incoming}
\eeq
where $V$ denotes a finite volume normalization of the two scattering states;
$E_{1,2}$ are the energies of the incoming and outgoing photons, respectively, while $\epsilon_{\alpha}(p_1)$ and $\epsilon^*_{\alpha}(p_2)$ are their polarization 
vectors. 
Using these notations, the same matrix element can be expressed in position space in terms of the operator $T^{\mu\nu}(x)$ on the incoming and outgoing photon eigenstates, multiplied by the corresponding incoming ($A^i$) and outgoing ($A^f$) wave functions
\beq
- i\, \frac{\kappa}{2}\, \langle p_2| T^{(0)\,\mu\nu}(x)|p_1\rangle = 
V^{\mu\nu\alpha	\beta}(p_1,p_2)A^i_\alpha(p_1)\, A^f_\beta(p_2) \, e^{i q\cdot x}.\, 
\eeq
We will be denoting with $q = p_2 - p_1$ the 4-momentum transfer on the graviton line. The scattering tree-level matrix element is then written as 
\bea
i\, \mathcal{S}^{(0)}_{if} &=& 
- i\, \frac{\kappa}{2}\, \int_{\mathcal V} d^4 x\, e^{i q\cdot x} h_{\mu\nu}(x) \hat{T}^{(0)\,\mu\nu}  \nn \\
&=&  2 \pi \delta(q_0) \, \mathcal{N}_i \, \mathcal{N}_f \, h_{\mu\nu}(\vec{q})\,
V^{\mu\nu\alpha\beta}(p_1,p_2)\, \epsilon_\alpha(p_1)\,\epsilon_\beta^{*}(p_2),  \, 
\eea
where ${\mathcal V}$ denotes the relevant region of integration allowed by the interaction.  
Using the conservation of the photon energy 
$(E_1=E_2\equiv E)$ and the elementary relation $p_1\cdot p_2 = E^2\, \left ( 1-\cos\theta \right)$, with $\theta$ the photon scattering angle, the square scattering matrix element takes the form
\beq
| i \mathcal{S}^{(0)}_{i f}|^2  = 
\left(-\frac{\kappa}{2}\right)^2 \, \left(\mathcal{N}_i \, \mathcal{N}_f \right)^2 \times (2\pi\delta(q_0)\mathcal T)
\times \left(\frac{\kappa M}{2 \vec{q}^2}\right)^2
\times
16\, E^4 \, \cos^4\left(\frac{\theta}{2}\right) \, ,
\label{averaged}
\eeq
where we have averaged over the polarizations of the initial photon and summed over the polarization of the final one.
As usual, we have extracted, from the square of the delta function, the transition time $\mathcal{T}$, 
using  $(2\pi\delta(q_0))^2=2\pi\delta(q_0)\mathcal{T}$.

To compute the differential cross section we multiply Eq.(\ref{averaged}) by the density of final states $d n_f$ given by 
\beq
d W = \frac{\vline\, i \mathcal{S}_{if}\,\vline^2}{j_i} dn_f,
\label{rate}
\eeq
normalized with respect to the incident photon flux $j_i=|\vec{p_1}|/(E_i V)$ in a volume $V$. We use the expression 
\beq
d n_f = \frac{V}{(2 \pi)^3} d^3 \vec{p}_f = \frac{V}{(2 \pi)^3}|\vec{p_2}|E_2 dE_2 d\Omega
\eeq
with $d\Omega= \sin\theta d\theta d\phi$ denoting the angular integration.
To obtain the differential cross section we need the transition rate per unit time $(d\sigma\equiv{d W}/{\mathcal{T} })$ that we integrate over the energy of the final photon to obtain the tree-level expression 
\beq
\frac{d \sigma}{d \Omega}_{0}=
(G\, M)^2 \cot^4\left(\frac{\theta}{2}\right)\, 
\label{leading}\
\eeq
which shares for small $\theta$ the typical $1/\theta^4$ behaviour of the Rutherford cross section. This expression can already be the starting point for a semiclassical analysis of the lensing of a photon in a weak gravitational background. 
Notice that  Eq.(\ref{leading}) is energy independent. Once this is inserted into (\ref{semic}) and solved for $b$ as a function of the angle of deflection, it allows to obtain the quantum counterpart of the classical lensing expression, which is also energy independent. We will come back to this point in the following sections. For the moment we prefer to move on with our analysis, by discussing the SM one-loop corrections to the Born level result 
given above. The link between the classical and the quantum Born level result will be re-addressed in section 7.

\section{One-loop corrections}
The analysis of the one-loop corrections to the interaction of the fields of the SM with gravity, in the weak gravitational field limit, has only received sporadic attention over several decades. In particular, the renormalizability 
of the SM Lagrangian in curved backgrounds, in the weak field limit, has never been considered in full generality, although specific results concerning this issue, at least for specific correlators, are available. For these reasons, in this section we are going to briefly elaborate on this point, which is crucial for our further analysis. We anticipate that our investigation will be restricted to the issue of renormalizability of the graviton/photon/photon interaction ($TAA$ vertex), with no claim to generality. 
As discussed in \cite{Coriano:2011zk}, the renormalizability is guaranteed if and only if the coupling of the Higgs doublet to the scalar curvature $R$ of the metric background is conformal.
Our analysis, in this case, is based on an explicit computation of the vertex, which is performed in the $R_\xi$ gauge and in dimensional regularization. In particular, the counterterms of the $TAA$ correlators are those derived from the SM Lagrangian at one-loop order.  \\
Originally, the radiative QED corrections to the propagation of fermions and photons were computed long ago by Berends and Gastmans in 
\cite{Berends:1975ah}, who showed the finiteness of the corresponding interactions.
The extension of those results to the SM case have been addressed only quite recently \cite{Degrassi:2008mw}, \cite{Coriano:2013iba, Coriano:2012cr,Coriano:2013msa}. The neutral currents sector, instead, has been discussed in \cite{Coriano:2011zk}.

\subsection{The vertex at one-loop order: overview}
\begin{figure}[t]
\centering
\subfigure[]{\includegraphics[scale=0.8]{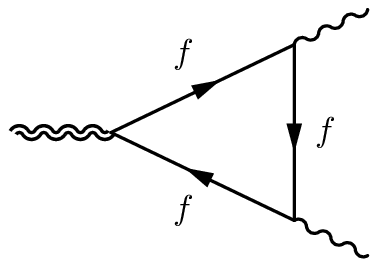}} \hspace{.5cm}
\subfigure[]{\includegraphics[scale=0.8]{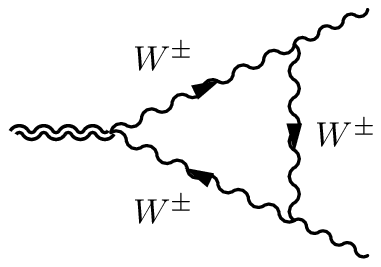}} \hspace{.5cm}
\subfigure[]{\includegraphics[scale=0.8]{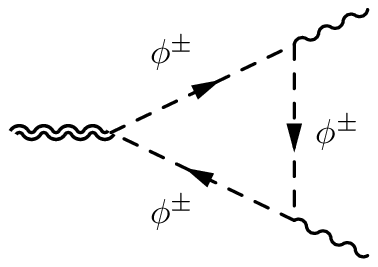}} \hspace{.5cm}
\subfigure[]{\includegraphics[scale=0.8]{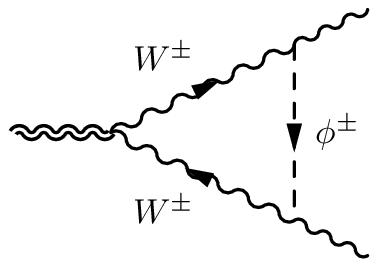}}
\\
\vspace{.5cm}
\subfigure[]{\includegraphics[scale=0.8]{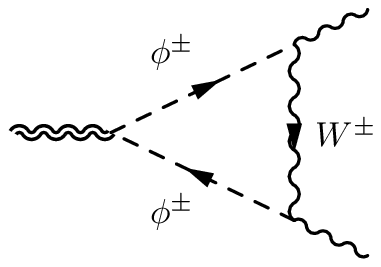}} \hspace{.5cm}
\subfigure[]{\includegraphics[scale=0.8]{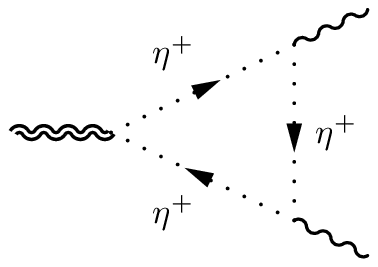}} \hspace{.5cm}
\subfigure[]{\includegraphics[scale=0.8]{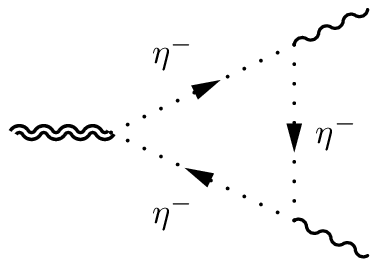}} \hspace{.5cm}
\caption{Amplitudes with the triangle topology.
\label{triangles}}
\end{figure}

\begin{figure}[t]
\centering
\subfigure[]{\includegraphics[scale=0.75]{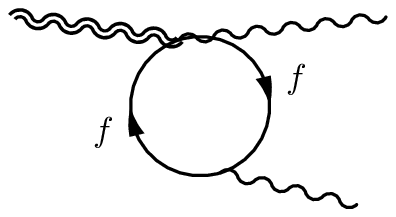}}\hspace{.5cm}
\subfigure[]{\includegraphics[scale=0.75]{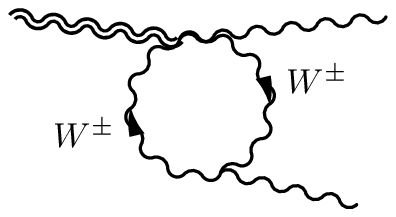}}\hspace{.5cm}
\subfigure[]{\includegraphics[scale=0.75]{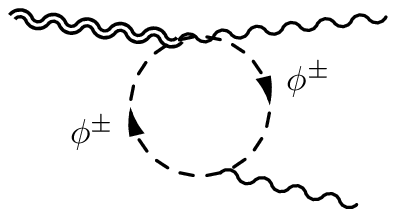}}\hspace{.5cm}
\subfigure[]{\includegraphics[scale=0.75]{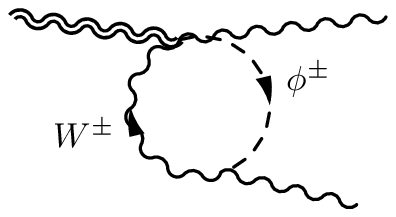}}
\\
\vspace{.5cm}
\subfigure[]{\includegraphics[scale=0.75]{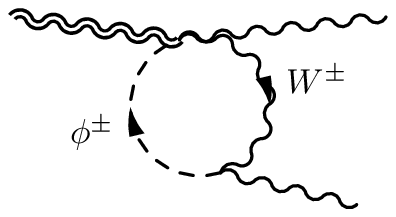}}\hspace{.5cm}
\subfigure[]{\includegraphics[scale=0.75]{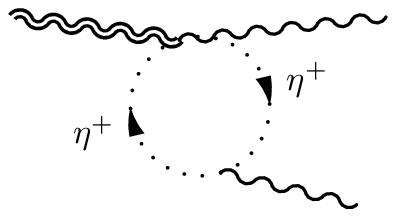}}\hspace{.5cm}
\subfigure[]{\includegraphics[scale=0.75]{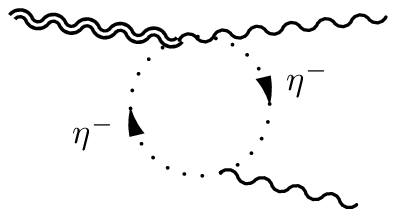}}\hspace{.5cm}
\caption{Amplitudes with t-bubble topology. \label{t-bubble}}
\end{figure}

\begin{figure}[t]
\centering
\subfigure[]{\includegraphics[scale=0.75]{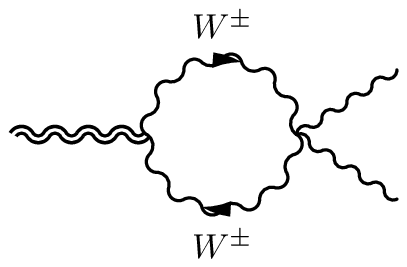}}\hspace{.5cm}
\subfigure[]{\includegraphics[scale=0.75]{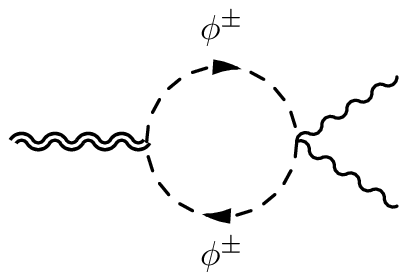}}
\caption{Amplitudes with s-bubble topology. \label{s-bubble}}
\end{figure}

\begin{figure}[t]
\centering
\subfigure[]{\includegraphics[scale=0.8]{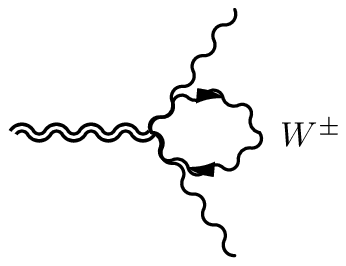}}\hspace{.5cm}
\subfigure[]{\includegraphics[scale=0.8]{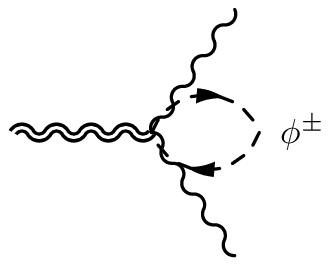}}
\caption{Amplitudes with the tadpole topology. \label{tadpoles}}
\end{figure}

The perturbative expansion at this order is characterized by several contributions with different topologies, that we are going to detail.  We show in Fig.\ref{triangles} the contributions of triangle-type, while in Fig.\ref{t-bubble} we have included the "t-bubble" diagrams, where the point of insertion of the EMT coincides with that of the gauge current. The diagrams characterized by two gauge bosons emerging from a single vertex are named "s-bubbles" and are shown in Fig.\ref{s-bubble}. Other contributions are distinguished by their tadpole topologies and are illustrated in Fig.\ref{tadpoles}. Obviously, all these terms are accompanied by similar diagrams with the exchange of the two photons. 

The complete one-loop correlator, that we call $\Gamma^{\mu\nu\alpha\beta}(p_1,p_2)$, can be organized into two contributions in the form
\beq \label{IncludeAmp}
\Gamma^{\mu\nu\alpha\beta}(p_1,p_2) = \Sigma^{\mu\nu\alpha\beta}(p_1,p_2) + \Delta^{\mu\nu\alpha\beta}(p_1,p_2)\, ,
\eeq
where $\Sigma^{\mu\nu\alpha\beta}(p_1,p_2)$ denotes the complete one-particle irreducible vertex, while $\Delta^{\mu\nu\alpha\beta}(p_1,p_2)$ is characterized by the presence of a bilinear graviton-Higgs mixing vertex on the external graviton line. As discussed in \cite{Coriano:2011zk}, the latter is necessary in order to consistently solve the Ward and the STI's  satisfied by the same correlator. 
The completely cut vertex, $\Sigma^{\mu\nu\alpha\beta}(p_1,p_2)$, on the other hand, can be split into the sum of three different sectors, depending on the particles running inside the loops, 
\bea
\Sigma ^{\mu\nu\alpha\beta}(p_1,p_2) = 
\Sigma_{F}^{\mu\nu\alpha\beta}(p_1,p_2) + 
\Sigma_{B}^{\mu\nu\alpha\beta}(p_1,p_2) + 
\Sigma_{I}^{\mu\nu\alpha\beta}(p_1,p_2). \, 
\label{Sigma-exp}
\eea
These correspond to the fermion (F) sector, Fig.\ref{triangles}(a) and Fig.\ref{t-bubble}(a), which also include the electromagnetic contribution; the weak sector (B), involving the charged gauge bosons $W^\pm$ together with their Goldstones and ghosts, Fig.\ref{triangles} (b)-(g), Fig.\ref{t-bubble} (b)-(g), Fig.\ref{s-bubble} and Fig.\ref{tadpoles}, and, 
finally, the sector of "improvement" ($I$). The latter contributions are given by the diagrams depicted in Fig.\ref{triangles} (c), 
(e) and Fig.\ref{s-bubble} (b), with the graviton - scalar - scalar vertices obtained from the $T_I^{\mu\nu}$.  As we have mentioned a suitable choice of $\chi$ ($\chi = 1/6$) in $T_I$ guarantees the conformal coupling of the Higgs.

 The contributions in Eq.(\ref{Sigma-exp}) can be expanded, for a given fermion of mass $m_f$, as
\bea
\Sigma ^{\mu\nu\alpha\beta}_{F}(p_1,p_2) &=&  \, 
\sum_{i=1}^{3} \Phi_{i\,F} (t,0, 0,m_f^2) \, \phi_i^{\mu\nu\alpha\beta}(p_1,p_2)\,, \\
\Sigma ^{\mu\nu\alpha\beta}_{B}(p_1,p_2) &=&  \, 
\sum_{i=1}^{3} \Phi_{i\,B} (t,0, 0,M_W^2) \, \phi_i^{\mu\nu\alpha\beta}(p_1,p_2)\,, \\
\Sigma ^{\mu\nu\alpha\beta}_{I}(p_1,p_2)
&=&  
\Phi_{1\,I} (t,0, 0,M_W^2) \, \phi_1^{\mu\nu\alpha\beta}(p_1,p_2) + 
\Phi_{4\,I} (t,0, 0,M_W^2) \, \phi_4^{\mu\nu\alpha\beta}(p_1,p_2) \,,
\eea
where the tensor basis is given by the following four tensors
\bea
\label{phis}
\phi_1^{\, \mu \nu \a \b} (p_1,p_2) 
&=&
(t \, \eta^{\mu\nu} - k^{\mu}k^{\nu}) \, u^{\a \b} (p_1,p_2)\, ,
 \label{widetilde1} \nn \\
\phi_2^{\, \mu \nu \a \b} (p_1,p_2) 
&=& 
- 2 \, u^{\a \b} (p_1,p_2) \left[ t \, \eta^{\mu \nu} + 
2 (p_1^\mu \, p_1^\nu + p_2^\mu \, p_2^\nu ) + 4 \, (p_1^\mu \, p_2^\nu + p_2^\mu \, p_1^\nu) \right] \, ,
\label{widetilde2} \nn \\
\phi^{\, \mu \nu \alpha \beta}_{3} (p_1,p_2) 
&=&
- \big(p_1^{\mu} p_2^{\nu} + p_1^{\nu} p_2^{\mu}\big)\eta^{\alpha\beta}
+ \frac{t}{2} \left(\eta^{\alpha\nu} \eta^{\beta\mu} + \eta^{\alpha\mu} \eta^{\beta\nu}\right) 
- \eta^{\mu\nu}  \, u^{\a \b} (p_1,p_2) \nn \\
&&
+ \left(\eta^{\beta\nu} p_1^{\mu}+ \eta^{\beta\mu} p_1^{\nu}\right)p_2^{\alpha}
 + \big (\eta^{\alpha\nu} p_2^{\mu} + \eta^{\alpha\mu} p_2^{\nu }\big)p_1^{\beta}\, , \nn \\
\phi_4^{\mu\nu\alpha\beta}(p,q) 
&=& 
(t \, \eta^{\mu\nu} - k^{\mu}k^{\nu}) \, \eta^{\alpha\beta}\, ,
\label{widetilde3}
\eea
with $u^{\a \b}(p_1,p_2)$ defined as
\beq
u^{\alpha\beta}(p_1,p_2) =   p_2^{\alpha} \, p_1^{\beta} - (p_1\cdot p_2) \,  \eta^{\alpha\beta}\, .
\label{utensor}
\eeq
The first three arguments of the form factors $\Phi$ in the expressions above stand for the three independent kinematical invariants,  
$k^2 = (p_2-p_1)^2 = t$, $p_1^2 = p_2^2 = 0$ respectively, while 
the remaining one denotes the masses of the particles circulating in the loop. 

We list below the radiative corrections coming from fermions. They would be the only contributions if we were considering only the abelian sector of the SM, i.e. QED. They are contained into 3 form factors for on-shell photons
\bea
\Phi_{1\, F} (t,\,0,\,0,\,m_f^2) 
&=& 
 \frac{\alpha}{3 \pi \, t} \, Q_f^2 \bigg\{
- \frac{2}{3} + \frac{4\,m_f^2}{t} - 2\,m_f^2 \, \mathcal C_0 (t, 0, 0, m_f^2, m_f^2, m_f^2)
\bigg[1 - \frac{4 m_f^2}{t}\bigg] \bigg\}\, ,  \nn \\
\Phi_{2\, F} (t,\,0,\,0,\,m_f^2)  
&=& 
 \frac{\alpha}{3 \pi \, t} \, Q_f^2 \bigg\{
-\frac{1}{12} - \frac{m_f^2}{t} - \frac{3\,m_f^2}{t} \mathcal D_0 (t, 0, 0, m_f^2, m_f^2)  \nn \\
&&  -\, m_f^2 \mathcal C_0(t, 0, 0, m_f^2, m_f^2, m_f^2 )\, \left[ 1 + \frac{2\,m_f^2}{t}\right] \bigg\}\, , \nn \\
\Phi_{3\,F} (t,\,0,\,0,\,m_f^2) 
&=&
 \frac{\alpha}{3 \pi \, t} \, Q_f^2 \bigg\{
\frac{11\,t}{12}+ 3 m_f^2 +  \mathcal D_0 (t, 0, 0, m_f^2, m_f^2)\left[5 m_f^2 + t \right]\nn\\
&&
+\, t \, \mathcal B_0 (0, m_f^2, m_f^2) + 3 \, m_f^2 \, \mathcal C_0(t, 0, 0, m_f^2 , m_f^2, m_f^2) \left[t + 2m_f^2 \right] \bigg\}\, ,
\eea
where we have included only the contribution due to a single fermion of mass $m_f$ and charge $Q_f$ running in the loops. We have denoted with $\alpha$ the fine structure constant. \\
The other gauge-invariant sector of the $TAA$ vertex is the one mediated by the exchange of the gauge bosons $W^\pm$ and of their corresponding Goldstones and ghosts. In this sector the form factors are explicitly given by the expressions
\bea
\Phi_{1\, B} (t,\,0,\,0,\,M_W^2) &=&  \frac{\alpha}{\pi \, t} \bigg\{
\frac{5}{6} - \frac{2\,M_W^2}{t} + 2\,M_W^2 \, \mathcal C_0 (t, 0, 0, M_W^2, M_W^2, M_W^2)
\bigg[1 - \frac{2 M_W^2}{t}\bigg] \bigg\},  
\eeqa
\bea
\Phi_{2\, B} (t,\,0,\,0,\,M_W^2)  &=& 
 \frac{\alpha}{\pi \, t} \bigg\{
\frac{1}{24} + \frac{M_W^2}{2\,t} + \frac{3\,M_W^2}{2\,t} \mathcal D_0 (t, 0, 0, M_W^2, M_W^2)  \nn \\
&+& 
\frac{M_W^2}{2} \mathcal C_0(t, 0, 0, M_W^2, M_W^2, M_W^2 )\, \left[ 1 + \frac{2\,M_W^2}{t}\right] \bigg\}\, , 
\eeqa
\bea
\Phi_{3\,B} (t,\,0,\,0,\,M_W^2) 
&=& 
 \frac{\alpha}{\pi \, t} \bigg\{
-\frac{15\,t}{8}-\frac{3\,M_W^2}{2} 
- \frac{1}{2}\, \mathcal D_0 (t, 0, 0, M_W^2, M_W^2)\left[5 M_W^2+\frac{7}{2} t \right]\nn\\
&-& 
 \frac{3}{4}t\, \mathcal B_0 (0, M_W^2, M_W^2)
- \mathcal C_0(t, 0, 0, M_W^2 , M_W^2, M_W^2) \left[t^2 + 4 M_W^2\,t + 3\,M_W^4\right]
\bigg\}. 
\eea
In the weak sector, we also have to consider the contributions related to the term of improvement, which couples the Higgs boson to the
graviton, as pointed out above in Eq.(\ref{thelagrangian}). 
The contributions due to this term are characterized by just two form factors
\bea
\Phi_{1\, I} (t,\,0,\,0,\,M_W^2) 
&=& 
 \frac{\alpha}{3 \pi \, t} \bigg\{ 1 + 2 M_W^2 \,C_0 (t, 0, 0, M_W^2, M_W^2, M_W^2)\bigg\} \,, \nn \\
\Phi_{4\, I} (t,\,0,\,0,\,M_W^2) 
&=&  
- \frac{\alpha}{6 \pi }  M_W^2 \,C_0 (t, 0, 0, M_W^2, M_W^2, M_W^2), \,
\label{improvff}
\eea
where we have already set the constant $\chi$ to 1/6.  
\begin{figure}[t]
\centering
\includegraphics[scale=0.8]{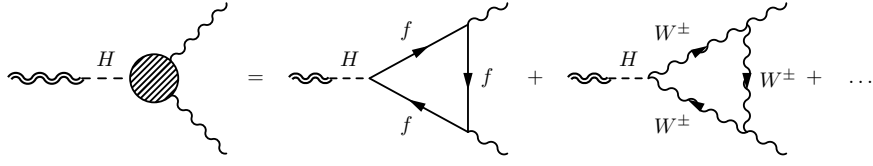}
\caption{Amplitude with the graviton-Higgs mixing vertex generated by the term of improvement. The blob represents the SM Higgs-photon-photon vertex at one-loop. \label{HVVImpr}}
\end{figure}
Finally, we also consider the correction to the external graviton leg, $\Delta^{\mu\nu\alpha\beta}(p_1,p_2)$,
to which the term of improvement contributes with the diagram depicted in Fig.\ref{HVVImpr}.
As discussed in \cite{Coriano:2011zk}, this contribution has to be included to solve consistently all the Ward and BRST identities
of the one-loop correlator. It is built by combining the tree-level vertex for graviton/Higgs mixing  - coming from the improved EMT -  with the
SM Higgs/photon/photon correlator at one-loop and it is given by
\bea
\Delta^{\mu\nu\alpha\beta}(p_1,p_2)  =  
\Psi_{1\, I} (t,0, 0,m_f^2,M_W^2,M_H^2) \, \phi_1^{\mu\nu\alpha\beta}(p_1,p_2) + 
\Psi_{4 \, I} (t,0, 0,M_W^2)  \, \phi_4^{\mu\nu\alpha\beta}(p_1,p_2)\, ,
\label{DAA}
\eea
where the on-shell form factors are 
\bea
\Psi_{1\, I}(t,\,0,\,0,\,m_f^2,M_W^2,M_H^2) &=& \Psi_{1\, I\,F}(t,\,0,\,0,\,m_f^2,M_H^2) + \Psi_{1\, I\,B}(t,\,0,\,0,\,M_W^2,M_H^2)\, ,  \nn \\
\Psi_{4\, I} (t,\,0,\,0,M_W^2) 
&=& 
- \Phi_{4\, I} (t,\,0,\,0,\,M_W^2), \, 
\label{ImprovementFF}
\eea
with
\bea
\Psi_{1\, I\,F}(t,\,0,\,0,\,m_f^2,M_H^2) &=& \frac{\alpha}{3 \pi \, t (t - M_H^2)}  2 m_f^2 \, Q_f^2 \bigg[ 2 + (4 m_f^2 -t) C_0 (t, 0, 0, m_f^2, m_f^2, m_f^2) \bigg] \nn \\
 \Psi_{1\, I\,B}(t,\,0,\,0,\,M_W^2,M_H^2)
&=&
 \frac{\alpha}{3 \pi \, t (t - M_H^2)} \bigg[ M_H^2 + 6 M_W^2  \nn \\
 && \qquad \quad+ 2 M_W^2 (M_H^2 + 6 M_W^2 - 4 t) C_0(t,0,0, M_W^2,M_W^2,M_W^2) \bigg] \,.
\eea
Notice that, given Eqs.(\ref{improvff}) and Eq.(\ref{ImprovementFF}), there is no contribution
to $\Gamma^{\mu\nu\alpha\beta}(p_1,p_2)$ coming from the $\phi_{4}^{\mu\nu\alpha\beta}(p_1,p_2)$ tensor structure. 

All the expressions presented above are given in terms of scalar integrals of two- and three-point functions, $\mathcal B_0$, $\mathcal D_0$ and $\mathcal C_0$, which have been defined as
\bea
\mathcal{B}_0(t, m^2,m^2)&=&
\frac{2}{\bar\epsilon} - \log\left(\frac{m^2}{\mu^2}\right)+ 2 
-  \tau(t,m^2) \, \log\left[\frac{\tau(t,m^2)+1}{ \tau(t,m^2)-1} \right] \, ,
\nn \\
\mathcal{B}_0(0, m^2,m^2)
&=& 
\frac{2}{\bar\epsilon} - \log\left(\frac{m^2}{\mu^2}\right) \, ,
\nn \\
\mathcal{D}_0(t,0, m^2,m^2)
&=& 
\mathcal{B}_0(t,m^2,m^2) -  \mathcal B_0(0,m^2,m^2) 
=
2 - \tau(t,m^2)\, \log\left[\frac{ \tau(t, m^2) +1}{ \tau(t,m^2) - 1}\right] \, ,
\nn \\
\mathcal{C}_0(t,0,0 ,m^2,m^2,m^2)
&=&
\frac{1}{2\,t}\, \log^2\left[\frac{ \tau(t,m^2) + 1 }{ \tau(t,m^2) -1}\right] \, , 
\eea
with
\beq
\tau(t, m^2)= \sqrt{1 - \frac{4\,m^2}{t}} \, .
\eeq
Here we have set $2/\bar\epsilon = 2/\epsilon - \gamma - \log \pi $, with $\epsilon=n-4$ in dimensional regularization in $n$ spacetime dimensions, and we have denotes with $\gamma$, as usual, the Euler-Mascheroni constant.
%
%

\subsection{The renormalization of the \texorpdfstring{$TAA$}{Lg} vertex}%
%
Concerning the ultraviolet behaviour of the previous form factors, we observe that only the $\mathcal{B}_0$ scalar integral is affected by a UV singularity. Therefore, the only UV divergent form factors which need a suitable renormalization prescription are $\Phi_{3\, F}$ and $\Phi_{3 \, B}$. All the other contributions are finite. \\
In our case, we have adopted the on-shell renormalization scheme in which the renormalization conditions are expressed in terms of the physical parameters of the theory. In the electroweak sector of the SM these are the masses of the physical particles, the electric charge and the quark mixing matrix. Moreover, we have required a unit residue of the 2-point functions on the physical particle poles, which define the wave function renormalization constants. These renormalization conditions allow to extract all the needed counterterms and are indeed sufficient to cancel all the UV singularities of the vertex functions built with an improved EMT on the external lines. We provide few more details on this point since the approach may not be so obvious. \\
From the counterterm Lagrangian of the SM one can compute the corresponding counterterm of the EMT, $\delta T^{\mu\nu}$, so that the bare $T^{\mu\nu}_0$ can be written in terms of the renormalized one $T^{\mu\nu}$ as
\bea
T^{\mu\nu}_0 = T^{\mu\nu} + \delta T^{\mu\nu} \,.
\eea
An explicit computation shows that the term of improvement in the EMT, in the conformally coupled case ($\chi = 1/6$), is necessary to ensure the finiteness of the complete one-loop 2-point function describing the bilinear mixing between the graviton and the Higgs scalar. \\
Concerning the $TAA$ correlator, the suitable counterterm vertex extracted from $\delta T^{\mu\nu}$ is given by
\bea
\label{TAAct}
\delta[TAA]^{\mu\nu\alpha\beta}(p_1,p_2) =  \bigg\{ - p_1 \cdot p_2 \, C^{\mu\nu\alpha \beta} + D^{\mu\nu\alpha \beta}(p_1,p_2) \bigg\} \delta Z_{AA} \,,
\eea
where $\delta Z_{AA}$ is the wave function renormalization constant of the photon field, defined as
\bea
\delta Z_{AA} = - \frac{\partial \Sigma^{AA}_T(k^2)}{\partial k^2} \bigg|_{k^2 = 0} \,,
\eea
with $\Sigma^{AA}_T$ being the transverse part of the one-loop photon self energy.
The photon wave function renormalization constant is explicitly given by
\beq
\delta Z_{AA} = \delta Z_{AA}^F + \delta Z_{AA}^B\, , 
\quad 
\delta Z_{AA}^F = \frac{\alpha\,Q_f^2}{3\,\pi}\, \left[  -\frac{2}{\overline{\epsilon}}  + \log\left( \frac{m_f^2}{\mu^2} \right) \right]\, ,
\quad
\delta Z_{AA}^B = \frac{\alpha}{2\,\pi}\, \left[  \frac{3}{\overline{\epsilon}} + \frac{1}{3}   
                                                                    - \frac{3}{2}\, \log\left( \frac{M_W^2}{\mu^2} \right) \right]\, ,
\eeq
where we have separated the contribution from the fermions from those due to the $W^{\pm}$ gauge bosons.\\
The tensor structure in Eq.(\ref{TAAct}), evaluated for on-shell photons, is exactly equal to $\phi_3^{\mu\nu\alpha\beta}(p_1,p_2)$, so that the counterterm only contributes to the UV-singular form factors, $\Phi_{3 \, F}$ and $\Phi_{3 \, B}$, as expected. \\%
The counterterm in Eq.(\ref{TAAct}) is sufficient to remove the divergence of the completely cut graph $\Sigma^{\mu\nu\alpha\beta}$ given in Eq.(\ref{Sigma-exp}). Furthermore, the $\Delta^{\mu\nu\alpha\beta}$ term in Eq.(\ref{DAA}), containing the bilinear mixing on the graviton line, is proportional to the SM Higgs/photon/photon vertex and does not require any renormalization, being finite. In more complicated cases such as, for instance, the $TAZ$ and the $TZZ$ correlators, $\Delta^{\mu\nu\alpha\beta}$ is instead UV divergent, but its finiteness is always guaranteed by the counterterms of the Lagrangian of the SM.\\
Notice that in the on-shell renormalization scheme there is no need to add the external leg corrections, or, equivalently, to consider the residue of the renormalized propagators on the external lines as prescribed by the LSZ formula. 
This is fixed to be 1 to all orders in perturbation theory by the renormalization conditions.
Therefore, we can combine the complete one-loop correction, including the counterterm, in the expression
\bea
\Gamma^{\mu\nu\alpha\beta}_{(1)}(p_1,p_2) 
&=&
- i\, \frac{\kappa}{2} \left(   
  \Sigma^{\mu\nu\alpha\beta}(p_1,p_2) + \Delta^{\mu\nu\alpha\beta}(p_1,p_2) + \delta Z_{AA} \, \phi_3^{\mu\nu\alpha\beta}(p_1,p_2)\right) 
\nn \\
& \equiv&
- i\, \frac{\kappa}{2}\, \sum_{i=1}^3 \phi_i^{\mu\nu\alpha\beta}(p_1,p_2)\, \overline{\Phi}_i \, ,
\label{IncludeAmp1}
\eea
where we have introduced a simplified notation for the UV finite form factors, $\overline{\Phi}_i = \overline{\Phi}_{i\, QED} + \overline{\Phi}_{i\, B}$, with
\bea
\overline{\Phi}_{1 \, QED} 
&=& 
\sum_f N_c^f\, \left( \Phi_{1\,F}(t,0,0,m_f^2) +  \Psi_{1\,I\,F}(t,0,0,m_f^2,M_H^2)  \right)  \, , \nn \\
\overline{\Phi}_{2 \, QED}
&=& 
\sum_f N_c^f \, \Phi_{2\,F}(t,0,0,m_f^2)  \, ,  \nn \\
\overline{\Phi}_{3 \, QED}
&=&
\sum_f N_c^f\, \left( \Phi_{3\,F}(t,0,0,m_f^2) + \delta Z_{AA}^F \right) \, , \nn \\
\overline{\Phi}_{1 \, B} 
&=&
\Phi_{1\,B}(t,0,0,M_W^2) + \Phi_{1\,I}(t,0,0,M_W^2)   +  \Psi_{1\,I\,B}(t,0,0,M_W^2,M_H^2) \,, \nn \\
\overline{\Phi}_{2 \, B}
&=&   \Phi_{2\,B}(t,0,0,M_W^2) \, ,  \nn \\
\overline{\Phi}_{3 \, B}
&=&
\Phi_{3\,B}(t,0,0,M_W^2) +   \delta Z_{AA}^B \, .
\label{IncludeAmp2}
\eea
In the expression above, the sum over $f$ is extended to all the charged fermions, while 
$N_c^f$ is a multiplicity factor, which is  1 for the leptons and  3 for the quarks, due to their color.

Notice also that all the form factors are UV finite and independent of the renormalization scale $\mu$, as one would expect in the on-shell renormalization scheme. An explicit computation shows that the same feature is recovered also in 
a mass-independent subtraction scheme such as dimensional regularization with modified minimal subtraction ($\overline{MS}$). In this case one can show that the one-loop corrections to the matrix element would assume the same form as in the on-shell scheme. In this case, the only difference would be in the definition of the masses and couplings, which in the $\overline{MS}$ are defined at the renormalization scale $\mu$ and do not coincide with the physical parameters present in the Lagrangian. Indeed, the differences in the expressions of the counterterms $\delta Z_{AA}$, evaluated in the two schemes, is exactly compensated by the addition of the external leg corrections. This peculiar feature is due to the intrinsic finiteness of the matrix element (defined as the sum of the vertex and of the external leg corrections) with a graviton on the external line, which is already manifest before the inclusion of any counterterm. In other words, if we had computed the $TAA$ matrix element by using the bare couplings, we would have obtained the same result. The on-shell scheme has been used both in the analysis of the UV finiteness of the cut vertex and in that of the complete vertex, with the inclusion of the external correction on the graviton leg.

\begin{figure}[t]
\centering
\includegraphics[scale=1.1]{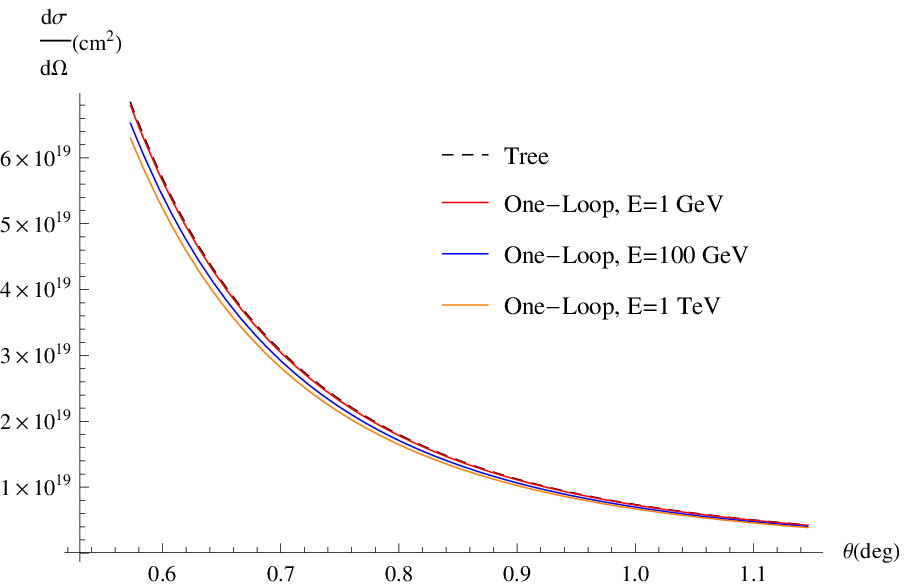}
\includegraphics[scale=1.1]{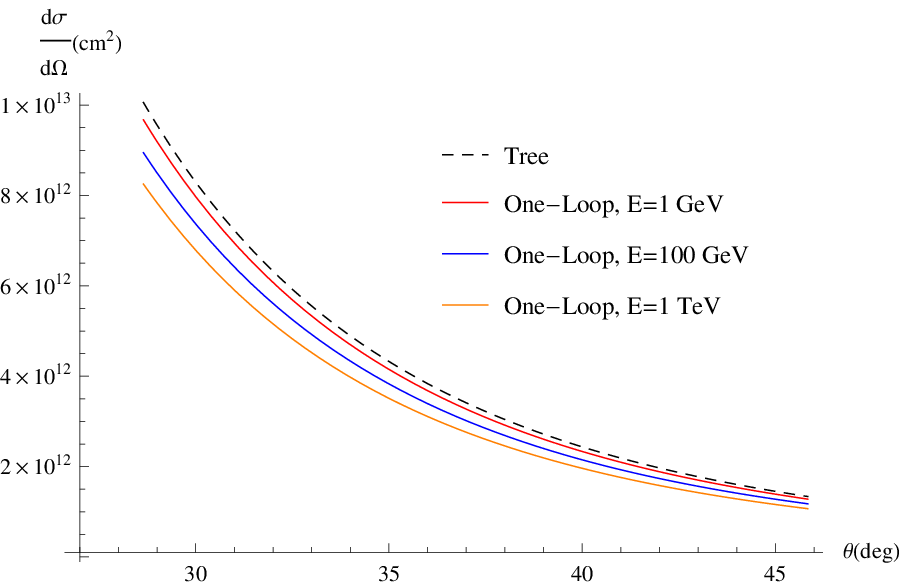}
\caption{Tree-level and one-loop cross sections for three values of the energy of the photon as a function of the scattering angles in two angular regions. }
\label{cs1}
\end{figure}

\begin{figure}[ht]
\centering
\includegraphics[scale=0.85]{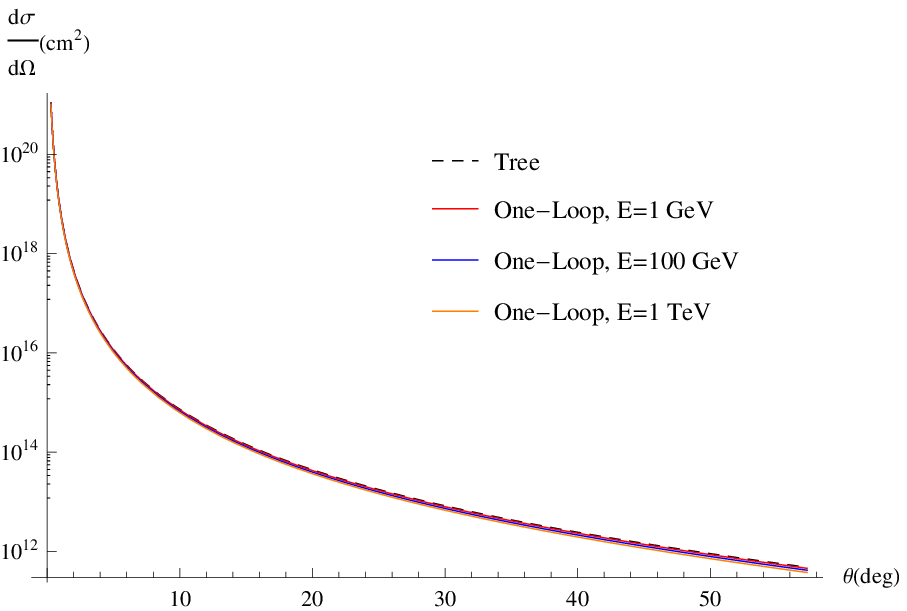}
\includegraphics[scale=0.85]{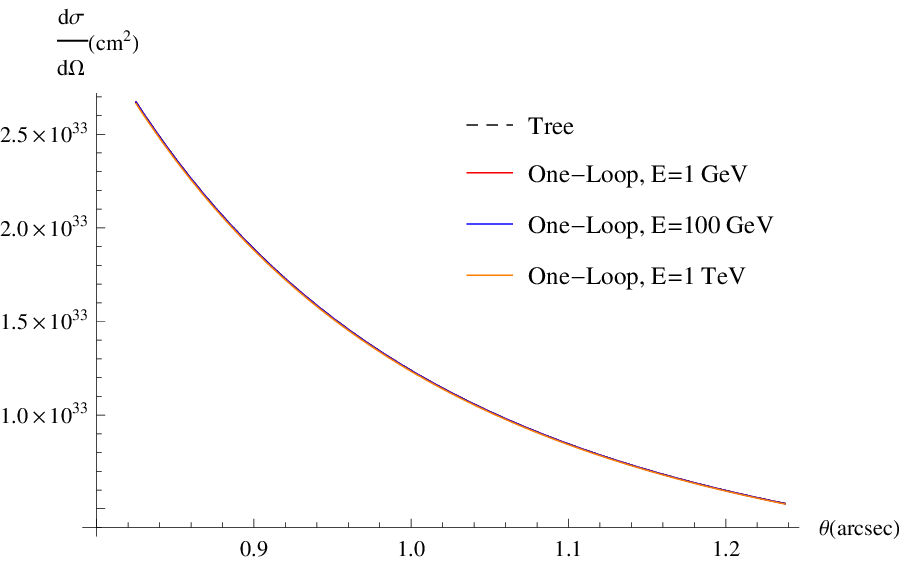}
\includegraphics[scale=0.90]{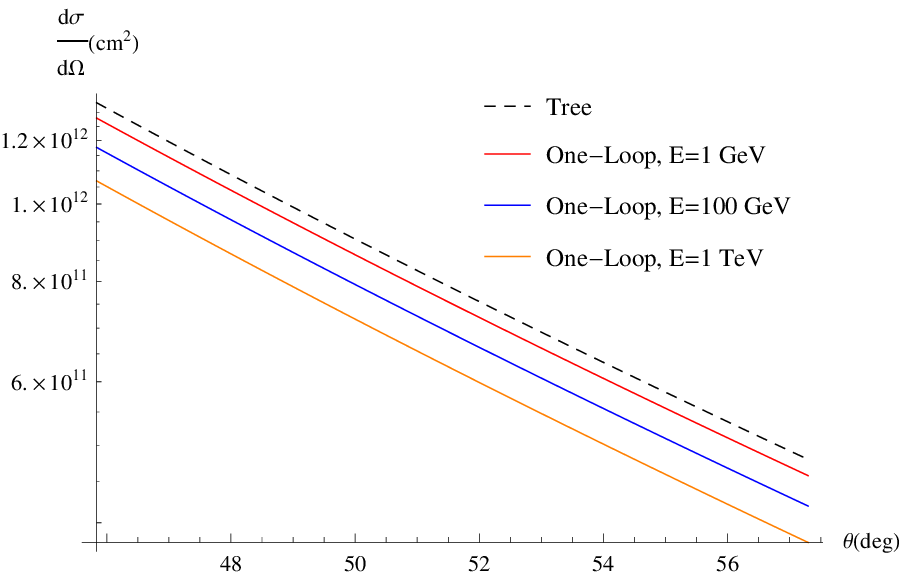}
\includegraphics[scale=0.90]{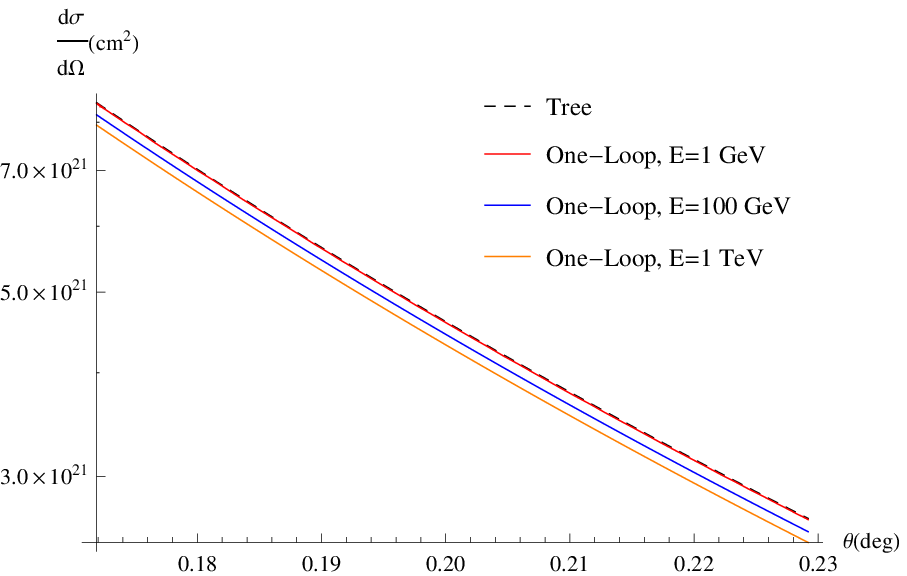}

\caption{Plots of the cross section as a function of the scattering angle for several energy values and for very strong and weak deflections (top). Zooms of two angular regions (bottom). }
\label{cs3cs4}
\end{figure}

\begin{figure}[t]
\centering
\includegraphics[scale=1.0]{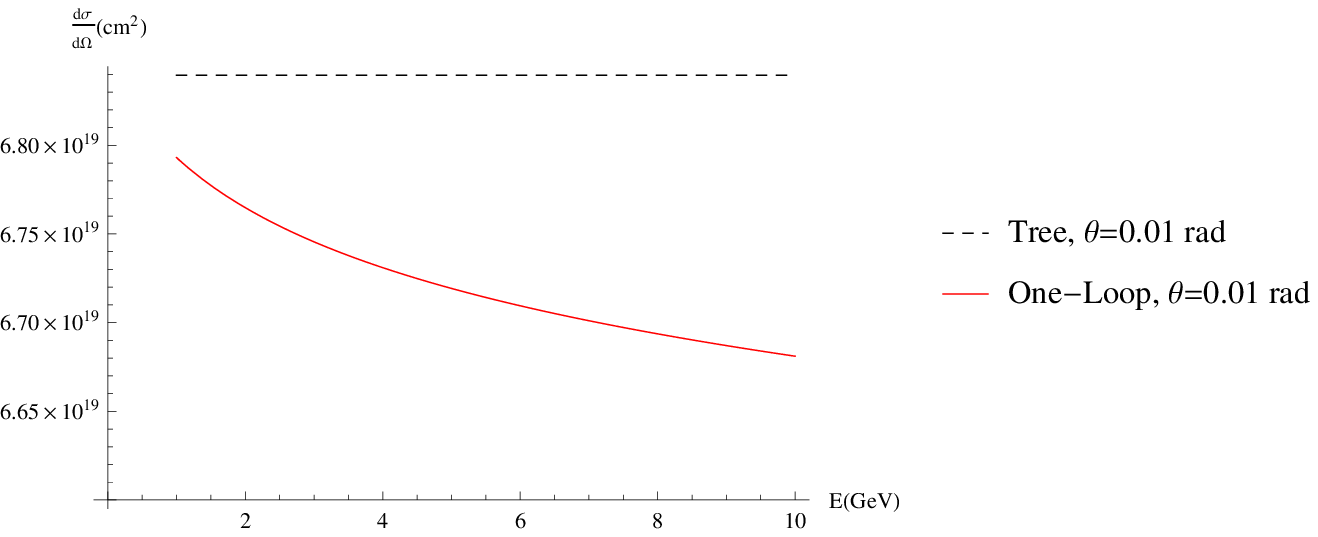}
\includegraphics[scale=1.0]{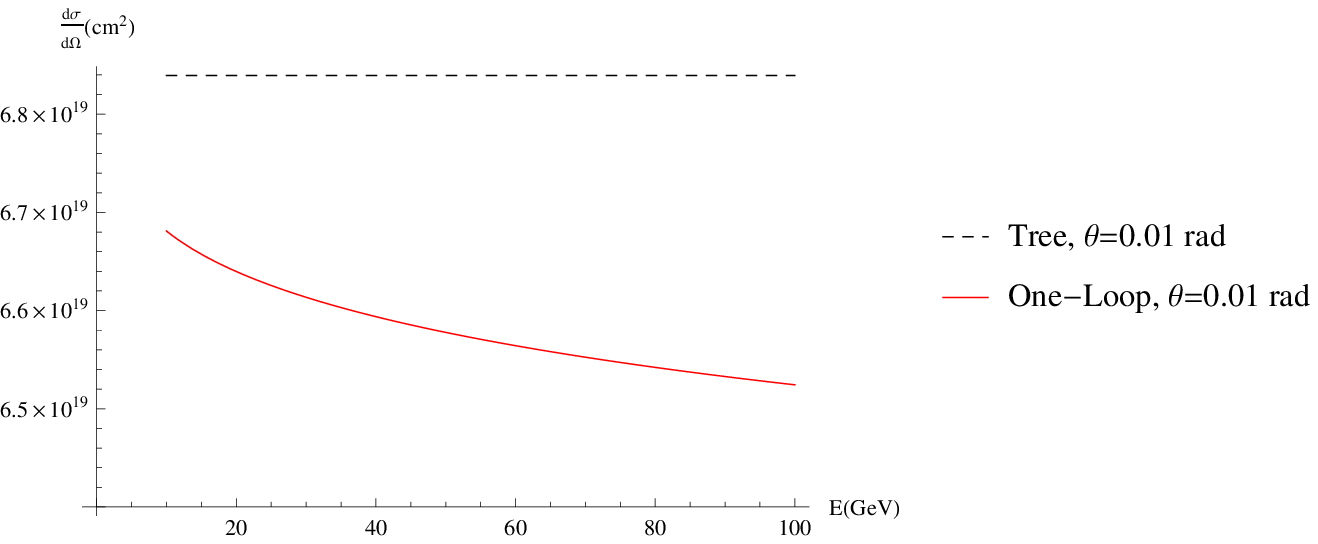}
\caption{Plots of the tree-level and one-loop cross sections as a function of the energy for a fixed scattering angle $\theta=0.01$ rad ($\theta= 0.57^{\circ}$).}
\label{cs6}
\end{figure}

\begin{figure}[t]
\centering
\includegraphics[scale=1]{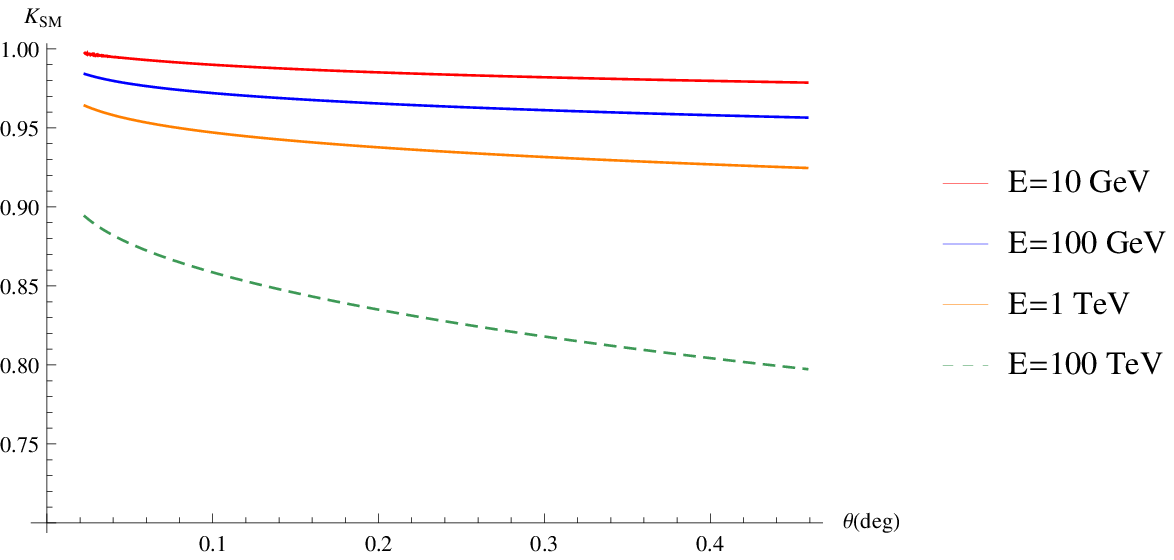}
\includegraphics[scale=1]{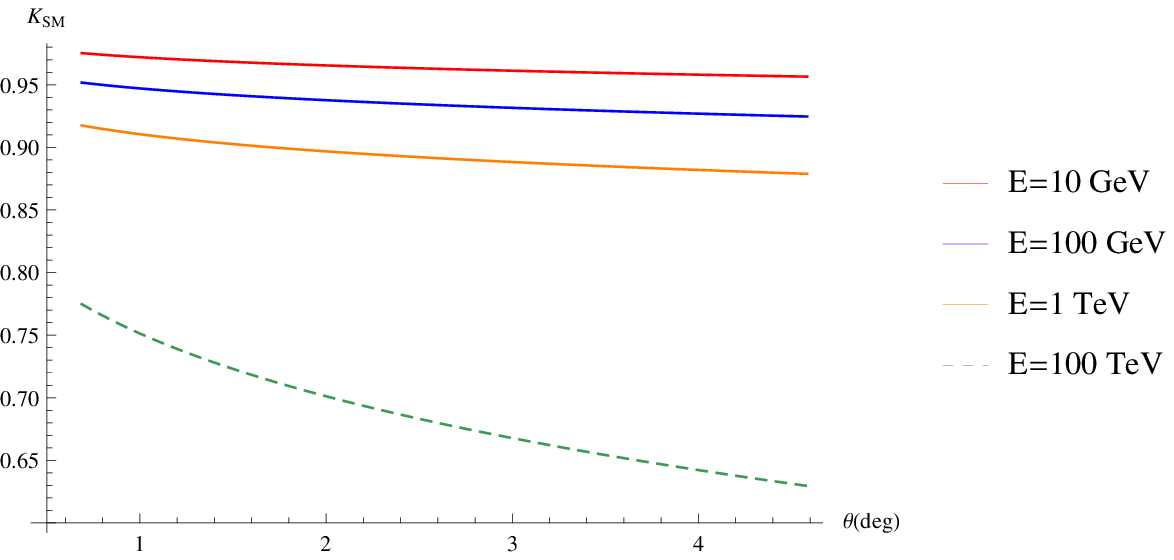}
\caption{$K_{SM}$ as a function of the scattering angle of the photon for four energy values, from the high energy to the ultra high energy (UHE) region.}
\label{kk1}
\end{figure}

\begin{figure}[t]
\centering
\includegraphics[scale=0.9]{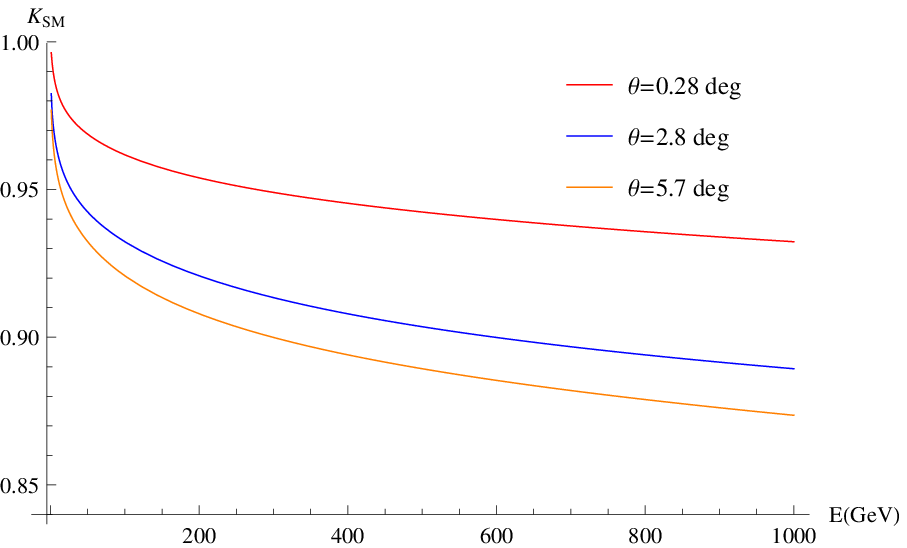}
\includegraphics[scale=0.8]{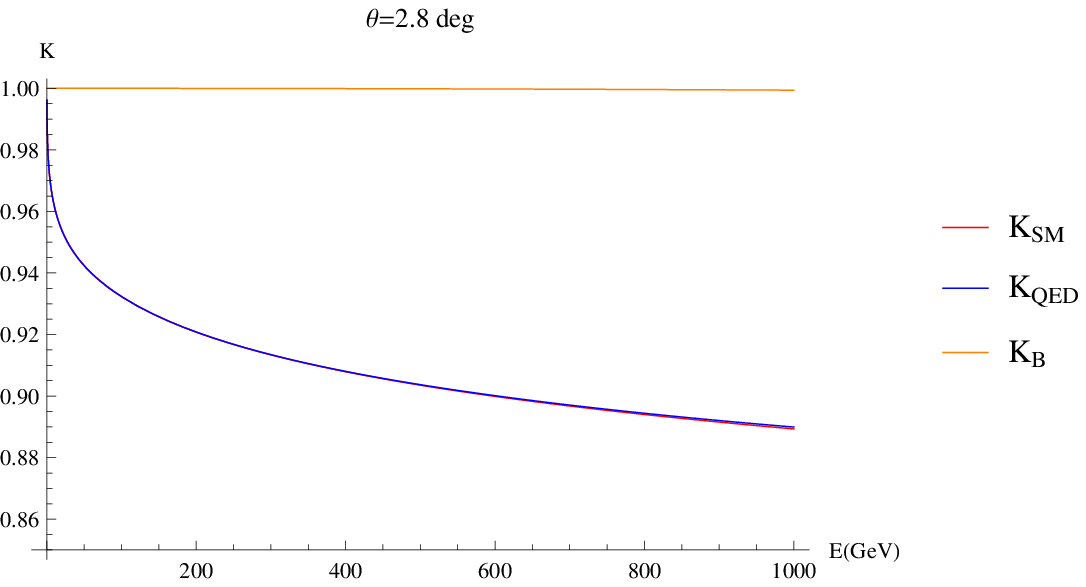}
\includegraphics[scale=0.9]{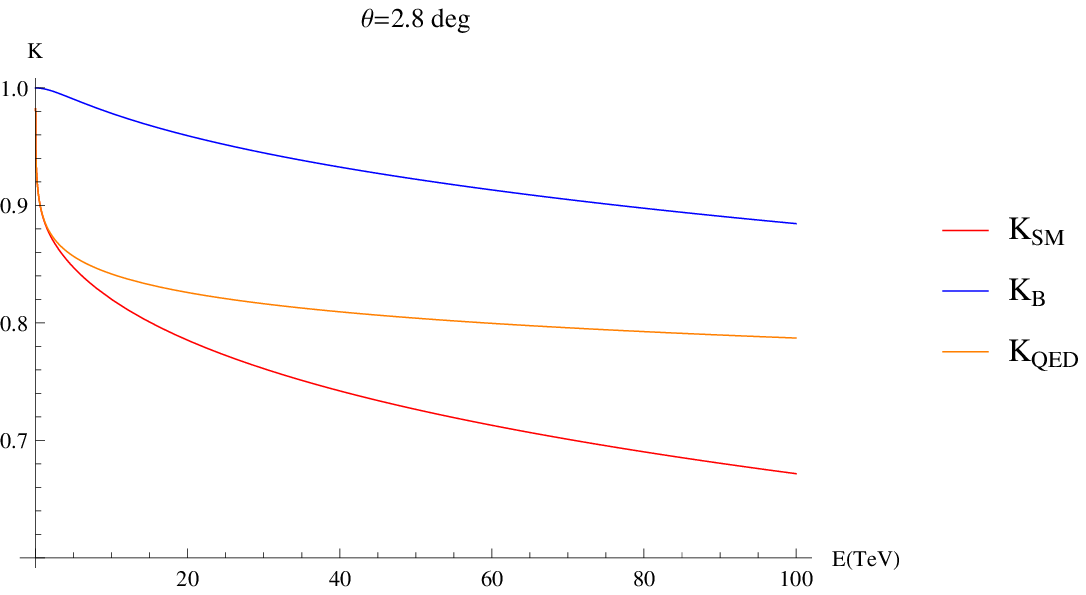}
\caption{A global plot of the $K_{SM}$-factor as a function of the energy for three values of the scattering angles (top-left). Plot of $K_{SM}$ and $K_{B}$ as functions of the photon energy below 1 TeV (top-right). Comparative plots of three form factors $K_{B}$, $K_{QED}$ and $K_{SM}$ as a function of energy up to 100 TeV (bottom). 
}
\label{kk2}
\end{figure}

\subsection{The cross section at next-to-leading order}%
We are now going to extend the computation of the cross section and include in our analysis the contributions coming from the one-loop vertex that we have just illustrated. \\
The matrix element at one-loop order can be defined as
\bea
i\, \mathcal S^{(1)}_{if} 
&=&
2\,\pi\, \delta(q_0)\, \left( \frac{\kappa\,M}{2\,|\vec{q}|^2} \right)\, \overline{S}_{\mu\nu}\, 
\left(  V^{\mu\nu\alpha\beta}(p_1,p_2) + \Gamma_{(1)}^{\mu\nu\alpha\beta}(p_1,p_2)\right) \,
A^i_\alpha(p_1)\, A^f_\beta(p_2) \,.
\label{MatEl}
\eea
As we are going to show, the contributions proportional to the tensors $\phi_1$ and $\phi_2$ 
drop identically in the unpolarized cross section. Notice that $\Phi_1$ is the form factor which carries the contribution from the conformal anomaly \cite{Coriano:2011zk}. \\
The expression for the SM one-loop cross section, in terms of the form factors 
$\overline{\Phi}_3$, is then given by 
\beq
\frac{d \sigma}{d\Omega} = 
(G M)^2\, \cot^4 \frac{\theta}{2}\,\left( 1 + 2 \, \textrm{Re} \, \overline{\Phi}_3 \right) \, .
\label{csSM}
\eeq
In the numerical analysis that will present in the next sections, we will also investigate the following $K$-factors    
\beq
K_{SM}= \frac{\frac{d\sigma}{d\Omega}}{\frac{d\sigma}{d\Omega}_{0}} \,, \qquad 
K_{QED}= \frac{ \frac{d\sigma}{d\Omega}_{0} + \frac{d\sigma}{d\Omega}_{ QED}}{ \frac{d\sigma}{d\Omega}_{0}} \,, \qquad
K_{B}= \frac{ \frac{d\sigma}{d\Omega}_{0} + \frac{d\sigma}{d\Omega}_{B}}{ \frac{d\sigma}{d\Omega}_{0}} \,, \qquad
\label{kfact}
\eeq
which will allow to quantify the size of the radiative corrections respect to the tree-level result. In this case, $K_{SM}$ quantifies the 
complete one-loop effects due to all the SM contributions, while $K_{QED}$ and $K_B$ account only for the QED and gauge boson exchanges respectively.  
    
In the previous equation, the partial contributions from the QED and the weak boson $(B)$ sector have been respectively defined as
\bea
\frac{d\sigma}{d\Omega}_{QED} &=&2 (G M)^2\left( \cot^4 \frac{\theta}{2}\right)  \textrm{Re} \, \overline{\Phi}_{3 \, QED}  \,,\label{csqed1f} \\
\frac{d\sigma}{d\Omega}_{B} &=&2 (G M)^2\left( \cot^4 \frac{\theta}{2}\right)  \textrm{Re} \, \overline{\Phi}_{3 \, B}  \,.
\eea
After taking into account the differences in the respective notations,  the QED the cross section in Eq.(\ref{csqed1f}) is in agreement with the result presented in \cite{Berends:1975ah}. 

\section{Photon scattering: numerical results}
In this section we will proceed with a numerical analysis of the results of the cross section and of the corresponding perturbative K-factors, before moving to a discussion of the differential equation of the impact parameter Eq.(\ref{semic}). For definiteness, we have chosen the mass of the gravitational source $M$ to be of the order ot the solar mass ($M=1.4 \,M_\odot$). \\
We show in Fig.\ref{cs1} the results for the tree-level and one-loop cross sections, for three values of the energy of the incoming photon, for 1 GeV, 100 GeV and 1 TeV respectively. The plots show that the impact of these corrections become  more sizeable as the energy increases. In general, they cause a reduction of the cross section, compared to the tree level result, over the entire angular interval. We have selected two different angular regions in  these plots. The intervals concern the small $\theta$ region, with angles around 1 degree, and the region of large $\theta$, which becomes significant only for scatterings very near to the horizon. It should be clear though, that due to the presence of a horizon,  which is not taken into account by our external metric, the behaviour obtained for this second region should not be considered predictive. In fact the gravitational field, as we approach the photon sphere, becomes strong, and the metric fluctuations that we consider in our analysis, described by Eq.(\ref{hh}) should be modified.
We will come back to a discussion of this important point in the second part of our work, when we turn to the semiclassical implications of these results. 
Notice also the strong suppression - approximately by 7 orders of magnitude - of the size of the cross section in the two regions.\\
The complete cross section as a function of $\theta$ is shown in Fig.\ref{cs3cs4} (top-left panel) for the same energies (1 GeV, 100 GeV and 1 TeV). We have compared the tree-level and the one-loop result, with the inclusion both of the QED and of the weak contributions. 
The divergence at small $\theta$ is obviously due to the $1/\theta^4$ behaviour generated by the prefactor wich accompanies both the tree-level and the one-loop cross sections. The top-right panel in the same figure is a zoom of the cross section in the region of small deflections, of the order of 1 arcsecond. We will see, from the semiclassical analysis of the next sections, that such deflections are obtained for scatterings with impact parameters about $10^6\, b_h$. We recall that $b_h$ is the impact parameter in horizon units, namely $b_h = b /(2 G M)$. Notice that in this panel the curves for different photon energies are completely overlapping. 
In two additional panels - reported in the same figure - we show the cross section both in the regions of very large (bottom-left panel, $\sim 50^\circ$) and of smaller angles (bottom-right panel, $\sim 0.2^\circ$). \\
Additional information on the size of the radiative contributions can be gathered from Fig.\ref{cs6}. Here we plot the cross section as a function of the photon energy, for a fixed angle $\theta=10^{-2}$ rad ($\theta=0.57^\circ$). 
 The results shows that such a dependence is quite mild over most of the high energy spectrum, from 1 GeV to 10 GeV (top panel) and from 10 GeV to 100 GeV (bottom panel), reaching approximately a correction of 4 $\%$ only in region of extremely high energy. \\
The impact of the radiative corrections and of the various contributions (QED, weak) to the scattering process emerges also from the plots of the various K-factors defined in Eq.(\ref{kfact}). For this purpose, we show in Figs.\ref{kk1} and \ref{kk2} the results for the K-factors both as a function of the scattering angle (Fig.\ref{kk1}), and of the energy (Fig.\ref{kk2}). In Fig.\ref{kk1} we have selected a scattering in an angular region with $\theta$ around $0.2^\circ$ and four different values of the photon energy (top panel). As the energy raises, the impact of the corrections is enhanced, from about the 4 $\%$ value, for scatterings in the range of several GeV' s, up to almost 20 $\%$ in the extremely high energy TeV region. In the bottom panel we repeat the analysis for larger scattering angles, using the same setup. It is clear that the radiative effects are more enhanced in this compared to the previous case, reaching about the value of $30 \%$ at extremely high energy. \\
In Fig.\ref{kk2} we present three additional panels where we show the variation of all the K-factors as functions of energy and for specific values of the scattering angles. In the top-left panel we plot the electroweak K-factor up to an energy of 1 TeV, for values of the scattering angles equal to 0.28, 2.8 and 5.7 degrees. The plot follows rather closely the trend already encountered in Fig.\ref{kk1}, evidencing the departure of the various K-factors from unity as the energy grows. In the top right panel we compare the QED $K_{QED}$, weak $K_B$ and $K_{SM}$ factors, plotted as a function of energy. From these plots it is evident that the weak corrections are extremely small, being the $K_B$ factor completely dominated by the tree-level contributions, which makes it equal to 1 over the entire range (top-right panel). The overwhelming part of the corrections, below 1 TeV, is therefore of QED origin, as expected. Finally, we show in the panel at the bottom of the same figure all the K-factors plotted versus the energy - up to an extremely high scale of 100 TeV - for a scattering angle of $\theta=2.8^\circ$. It is clear that both the weak 
($K_B$ contributions) and the QED sectors ($K_{QED}$) become significant at extremely high energy. The weak contributions amount to $10 \%$ of the variation of the $SM$ cross section in the asymptotic limit. The QED part is responsible for approximately a $20 \%$ reduction of the same cross section, with a $K_{SM}$  around $0.7$ at 100 TeV.

\subsection{Polarization }\label{Helicity}
Radiative corrections of the graviton/photon/photon vertex induce a helicity flip amplitude which contributes to the change in the polarization of a photon beam and as such requires a special attention. In this section we proceed with a discussion of these effects.\\ 
We assume that the incoming photon carries a momentum $p_1$ oriented along the $z$ axis and that the
scattering takes place in the $z-y$ plane, so that the two momenta can be parameterized as
\beq
p_1^\mu = E\, (1,0,0,1)\, , \quad p_2^\mu = E\, (1,0, \sin\theta , \cos\theta ) \, .
\label{momenta}
\eeq
Correspondingly, the polarization vectors are
\bea
\epsilon^{\mu\,\pm}(p_1) &=& \frac{1}{\sqrt{2}}(0,1,\pm i,0) \nn \\
\epsilon^{\mu\,\pm}(p_2) &=& \frac{1}{\sqrt{2}}(0,1,\mp i\, \cos\theta,\pm i\, \sin\theta).
\label{polarization}
\eea
Starting from the matrix element defined in Eq.(\ref{MatEl}),  using the definition of the photon plane waves in Eq.(\ref{incoming}),
Eq.(\ref{momenta}) and Eq.(\ref{polarization}), and adopting the notations of Eq.(\ref{IncludeAmp1})-(\ref{IncludeAmp2}), we obtain for the helicity amplitudes the expressions
\bea
i\, \mathcal{S}^{++}_{if} 
&=& 
i\,8\p^2\, \delta(q_0)\, \mathcal{N}_i\, \mathcal{N}_f\,G\, M \, \cot^2\left(\frac{\theta}{2}\right)\,
\bigg( 1 - \overline{\Phi}_3  \bigg) \, ,
\nn \\
i\, \mathcal{S}^{+-}_{if} 
&=&
i\,16\p^2\, \delta(q_0)\, \mathcal{N}_i\, \mathcal{N}_f\,G\, M \, E^2\, 
\bigg(  8\, \overline{\Phi}_2 - \overline{\Phi}_1 +  \left( \overline{\Phi}_1 + 4\, \overline{\Phi}_2 \right)\cos^2\left(\frac{\theta}{2}\right) \bigg).
\eea
\begin{figure}[t]
\centering
\includegraphics[scale=0.6]{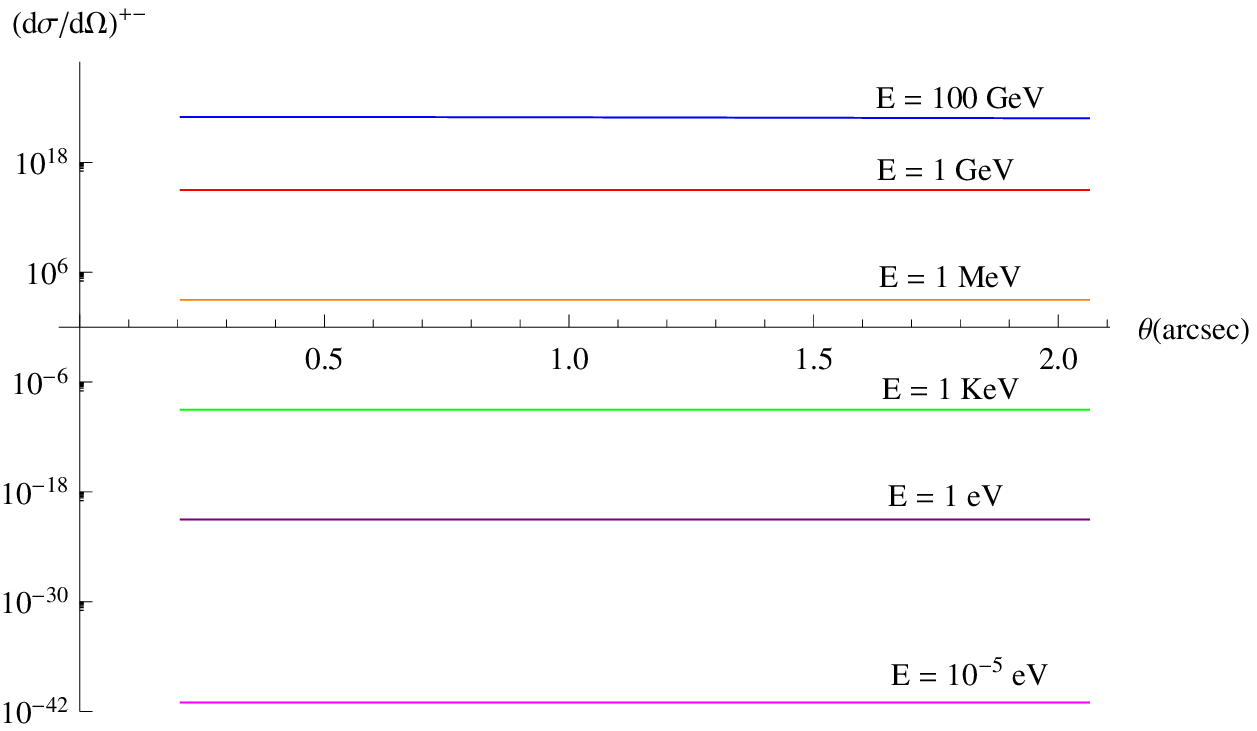}
\includegraphics[scale=0.9]{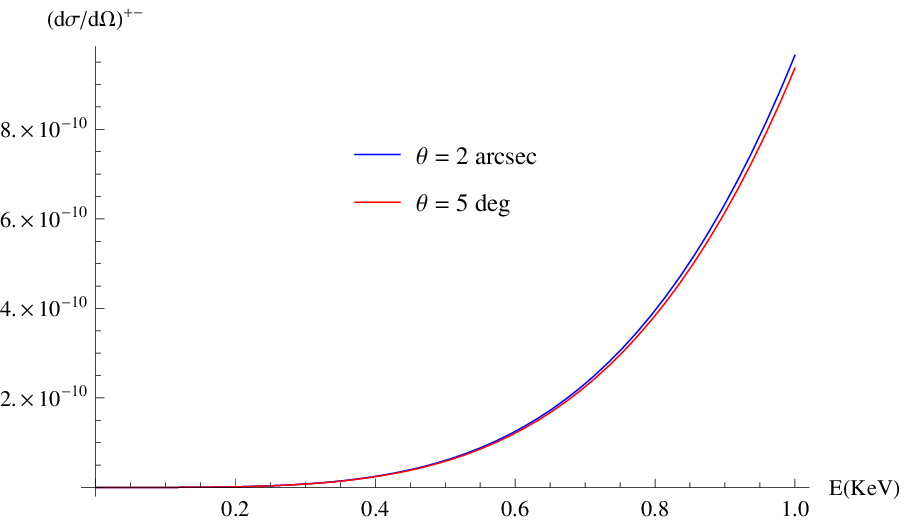}
\includegraphics[scale=0.9]{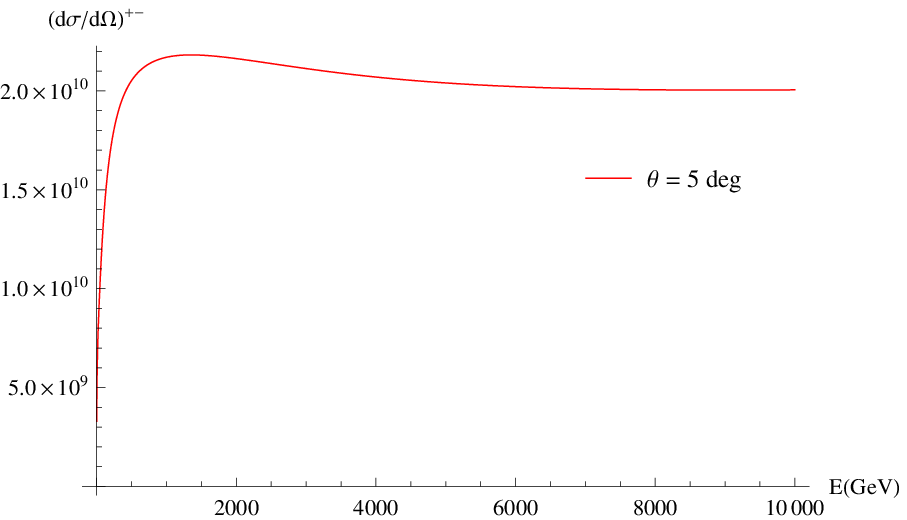}
\includegraphics[scale=0.9]{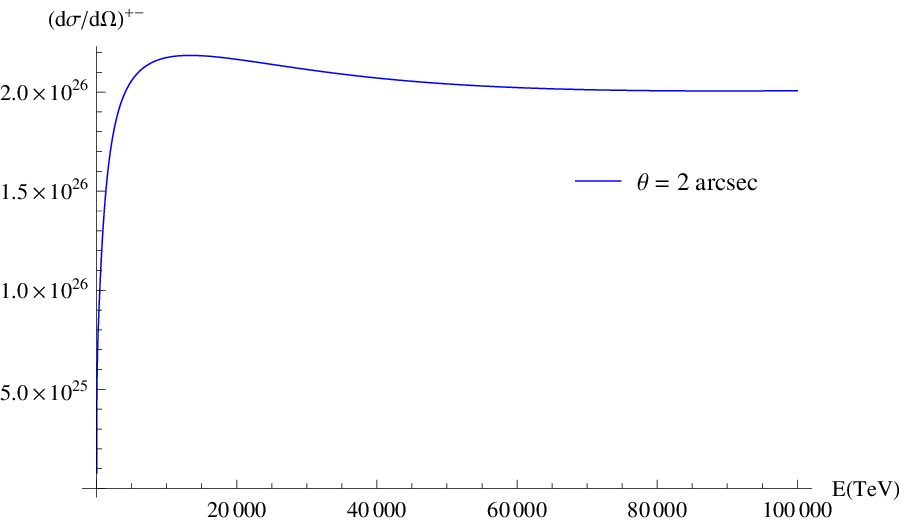}
\caption{   $(d\s/d\Omega)^{+-}$ as a function of the scattering angle for several values of the energy (top-left). $(d\s/d\Omega)^{+-}$ as a function of the photon energy for $\q = 2$ arcsec  and   $\q = 5$ deg (top-right). $(d\s/d\Omega)^{+-}$ as a function of the photon energy up to $10^4$ GeV for $\q = 5$ deg (bottom-left) and up to $10^5$ TeV for $\q=2$ arcsec (bottom-right).}
\label{pm1}
\end{figure}
Notice that the form factor $\overline{\Phi}_1$, which is associated to the trace anomaly in the massless case, appears in the helicity-flip amplitude but is does not contribute to the unpolarized
cross section. We obtain 
\bea
\frac{d \sigma}{d \Omega}^{++} &=& \frac{d \sigma}{d \Omega}^{--} = (G M)^2 \cot^4 \left( \frac{\theta}{2} \right) \left( 1 - 2 \, \textrm{Re} \, \overline \Phi_3\right) \,, \\
\frac{d \sigma}{d \Omega}^{+-} &=& \frac{d \sigma}{d \Omega}^{-+} = 4 (G M)^2 E^4 \bigg| 8 \overline \Phi_2 - \overline \Phi_1 + (\overline \Phi_1 + 4 \overline \Phi_2) \cos^2 \left( \frac{\theta}{2} \right)  \bigg|^2 \,.
\label{polplusminus}
\eea
Notice that the $(++)$ contribution coincides with the unpolarized one 
\bea
\frac{d \sigma}{d \Omega}&=&\frac{1}{2}\left( \frac{d \sigma}{d \Omega}^{++} +  \frac{d \sigma}{d \Omega}^{--}\right)=  \frac{d \sigma}{d \Omega}^{++}, 
\label{firstorder}
\eeqa
being only the $(++)$ cross section of $O(\alpha)$, while the $(+-)$ one is, respectively, of 
 $O(\alpha^2)$. The numerical analysis shows that the helicity-flip $(+ -)$ differential cross section for photons of the CMB is very small and 
 it is equal to $9.6\times 10^{-42}$ cm$^2$ over the entire angular range ($M=1.4 \,M_\odot$) with variations which are around one per thousands. For a very massive black hole of a million solar masses the cross section raises to $10^{-30}$ cm$^2$, but it is still far smaller than the corresponding Thomson cross section ($10^{-24}$ cm$^2$) describing the scattering of a photon off a single electron.\\
 We show in Fig.\ref{pm1} (top-left panel) a plot of the $(+-)$ cross section as a function of $\theta$, for $\theta$ about 1 arcsecond, for several values of the energy. The plot indicates that the cross section changes very significantly with the energy. For instance, in the CMB case the size of this helicity-flip contribution is negligible, as pointed out above, but at 100 GeV, for instance, it grows to $10^{22}$ cm$^2$. This growth however, is not sufficient to identify this contribution as a significant component of the unpolarized cross section, since the helicity-conserving $(++)$ one is by far the most dominant part, being about 10$^{33}$ cm$^2$ (see Fig.\ref{cs3cs4} top-right panel). \\
The plots of the $(+-)$ cross section as a function of the photon energy, in the case of weak ($\theta=2$ arcsec) and strong lensing ($\theta=5$ deg), are depicted in the top-right and bottom (left and right) panels. They show that the cross section starts from zero and grows with the energy as $\sim E^4$ for small $E$ (top-right panel). It reaches a maximum at a value of the energy which depends on the angle, and then stabilizes at higher energies, with a final plateau. The maximum is obtained for a momentum transfer $t$ of the order of the square of the top quark mass. 

\section{Photon lensing: Classical and semiclassical results}
\begin{figure}[t]
\centering
\includegraphics[scale=0.60]{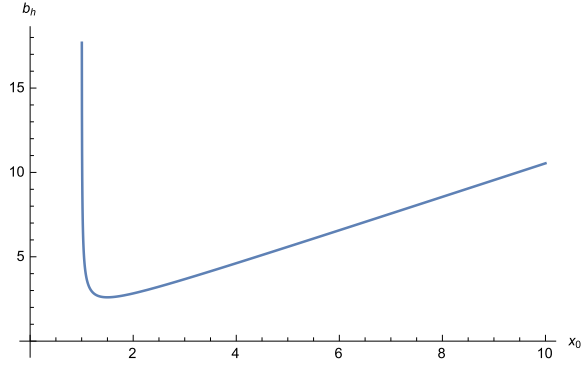}
\caption{ Plot of $b_h$ versus $x_0$, showing the singularity at the position of the photon sphere for $x_0=3/2$.  } 
\label{ddd}
\end{figure}

In this section we start investigating the semiclassical approach to the evaluation of the angular deflection of a photon using the notion of impact parameter, as expressed by Eq.(\ref{semic}). For this purpose, we briefly summarize the classical GR result concerning the deflection of light in a Schwarzschild background, presenting along the way some new numerical results which we deem necessary in order to proceed with our analysis. In this case the metric is given by Eq. (\ref{ds0}) with $C= 2 G M$, and the equations of motion along a null geodesic are 
\beq
g_{\mu\nu}\frac{d x^\mu}{d \lambda} \frac{d x^\nu}{d \lambda}=0,
\eeq
with $\lambda$ an affine parameter of the geodesic.
The equations of motion can be separated in the form
\bea
\left(1-\frac{2 M}{ r}\right)\frac{d t}{d\lambda} = E \qquad 
r^2 \frac{d\phi}{d\lambda}=J \qquad 
\frac{d \theta}{d\lambda}=0,
\eeqa
corresponding to the energy $(E)$ and the angular momentum $(J)$ of the massless particle. By setting 
\beq
u\equiv \frac{J}{E} 
\eeq
the geodesic equation becomes 
\beq
\left(1- \frac{2 M}{r}\right)\frac{1}{r^2} +\frac{1}{J^2} \left(\frac{d r}{d\lambda}\right)^2 -\frac{1}{u^2}=0,
\label{ge}
\eeq
with $u$ denoting the impact parameter $(u\equiv b_h)$.
The angle of deflection is expressed in the form 
\beq
\theta_d=2\int_{r_0}^\infty \frac{ d\phi}{d r } \, dr- \pi,
\eeq
where $r_0$ is the distance of closest approach between the photon beam and the lens. The subtraction of $\pi$ allows to remove the angular displacement in the absence of lensing. Using Eq.(\ref{ge}) we can re-express the deflection in the simplified form 
\beq
\theta_d(r_0)=\int_{r_0}^\infty d r \frac{2}{r^2} \left[  \frac{1}{u^2} -\frac{1}{r^2}\left(1 -\frac{2 M}{r}\right)\right] ^{-1/2} -\pi.
\eeq
The condition that $r_0$ is the point of closest radial approach between the source and the beam implies the extremum condition $d{r}/d\lambda=0$ which gives, from Eq.(\ref{ge}),
\beq
u=r_0\left( 1- \frac{2 M}{r_0}\right)^{-1/2}
\label{ufromr0}
\eeq
and thus
\beq
\theta_d(r_0)=\int_{r_0}^\infty dr \frac{2}{r^2}\left[ 
\frac{1}{r_0^2}\left(1 -\frac{2 M}{r_0}\right) -
\frac{1}{r^2}\left(1 -\frac{2 M}{r}\right)\right]^{-1/2} -\pi.
\label{exactT}
\eeq
The integral does not depend explicitly on the mass of the source, as far as we measure the distances in horizon units, $b_H\equiv b/R_S$, with the horizon given by the Schwarzschild radius $R_S=2 \,G M$. In fact, a simple rescaling of Eq.(\ref{ufromr0}) with $x_0\equiv r_0/(2 M)$  
gives
\beq
b_h \equiv u =x_0\left(1-\frac{1}{x_0}\right)^{-1/2},
\label{xb}
\eeq
and
\beq
\theta_d(x_0)=2\int_{x_0}^{\infty}\frac{dx}{x \sqrt{\left(\frac{x}{x_0}\right)^2\left(1- \frac{1}{x_0}\right) -\left( 1- \frac{1}{x}\right)}}-\pi.
\label{exactscaled}
\eeq
The expression above is not yet in the most appropriate form for a numerical analysis. An equivalent form is  
\beq
\theta_d(x_0) =\int_0^{1/x_0}\frac{2 \, dx}{\sqrt{\frac{1}{x_0^2}(1-\frac{1}{x_0})- x^2(1-{x})}} -\pi,
\label{erel}
\eeq
which can be used for a brute force numerical integration. The numerical results are as accurate as those obtained from the explicit expression given in terms of elliptic integrals. We will also be needing a numerical inversion of the function 
$\theta_d(x_0)$, being monotonic, thereby determining both $x_0$ and $b_h$ as functions of 
the angle of deflection. \\
It is also convenient to solve explicitly Eq.(\ref{xb}), which is important in the case of 
very strong lensing. We find that the relation between $x_0$ and $b_h$ can be inverted in the form 
\beq
x_0=\frac{\sqrt[3]{\frac{2}{3}} b_h^2}{\sqrt[3]{\sqrt{3} \sqrt{27 b_h^4-4 b_h^6}-9 b_h^2}}+\frac{\sqrt[3]{\sqrt{3}
   \sqrt{27 b_h^4-4 b_h^6}-9 b_h^2}}{\sqrt[3]{2} 3^{2/3}}.
\eeq
A plot of this relation is shown in Fig.\ref{ddd}, with a singularity located at $b_h=3/2 \sqrt{3}\equiv b_h^0$ (i.e. $x_0=3/2$), which is the point at which the angle of deflection in the GR expression Eq.(\ref{erel}) diverges.

An explicit evalutation of the integral \cite{Bozza:2001xd} gives 
\beq
\theta_d(x_0)=-\pi - 4 \, {\bf F}\left(\phi(x_0),\lambda(x_0)\right) \Sigma(x_0)
\label{deflformula}
\eeq
with 
 \beq
{\bf  F}(\phi,\lambda)=\int_0^\phi \frac{d\theta}{\sqrt{1- \lambda^2 \sin^2\theta}}
\eeq
being the elliptic integral of first kind, having defined, for simplicity, the auxiliary functions  
\bea
\phi_0(x_0) &=& \textrm{Arcsin}(\tau(x_0)) \,, \nonumber \\
\tau(x_0) &=&\sqrt{\frac{-3 + x_0 - \sqrt{-3 + 2 x_0 + x_0^2}}{2 (-3 + 2 x_0)}} \,, \nn\\
\Sigma(x_0)&=&\sqrt{\frac{{x_0} \left(-\sqrt{{x_0}^2+2 {x_0}-3}+3 {x_0}-3\right)}{(3-2 {x_0})
   \left(\sqrt{{x_0}^2+2 {x_0}-3}-{x_0}+1\right)}},
   \eea
 while 
 \beq
 \lambda(x_0)=\frac{  3 - x_0 -\sqrt{-3 + 2 x_0 + x_0^2}}{3 - x_0 +\sqrt{-3 + 2 x_0+x_0^2}}.
 \eeq
 Notice that in the equation above, both $\Sigma(x_0)$ and $\tau(x_0)$ are imaginary for 
 $x_0 > 3/2$, with the product of the elliptic integral ${\bf F}$ with $\Sigma$ being real. An alternative formula, 
 in terms of real factors is 
 \beq
\theta_d(x_0)=  
4 \sqrt{\frac{{2\, x_0}}{Y}} \left[{\bf F}\left(\frac{\pi}{2},\kappa\right)-{\bf F}\left(\textrm{Arcsin}\left(\sqrt{2} 
\sqrt{\frac{2\, {x_0}-2}{6\, {x_0}+Y-6}}\right),\kappa\right)\right]-\pi
 \label{pam}
 \eeq 
 with 
 \bea
 Y =\sqrt{4 ({x_0}-1) ({x_0}+3)} \,, \qquad 
\kappa = \frac{-{2\, x_0}+Y+6}{2Y}.
 \eea
 The nature of the singularity around $x_0=3/2$ can be easily worked out from Eq.(\ref{pam}) by setting 
$x_0=3/2 + \epsilon$  $(\epsilon \ll 1)$ and expanding the resulting expression to $O(\epsilon)$. One obtains 
\bea
\theta_d(3/2 + \epsilon)&\sim& -4{\bf  F}\left(\textrm{Arcsin}\left(\frac{1}{\sqrt{3}}\right),1\right)-\pi +\log
   (324) - 2 \log\epsilon \nonumber \nonumber \\
   &=& 0.00523507 - 2 \log\epsilon
   \eeqa 
 which proves to be logarithmically divergent as the beam approaches the photon sphere ($\epsilon\to 0$). \\
 The weak field expansion,valid for $x_0 \gg 3/2$, obtained from the elliptic solution, takes the form 
 \beq
 \theta_d(x_0)=\frac{2}{x_0} +\left(-1+\frac{15}{16}\pi\right)\frac{1}{x_0^2} + O(1/x_0^3).
 \eeq
 The first term in the expression above,  $\theta_d\sim 4 G M/b$, is Einstein's result, obtained in the case of a weak field, which for a photon skimming the sun is about 1.74 arcseconds. 
 
\begin{figure}[t]
\centering
\includegraphics[scale=0.7]{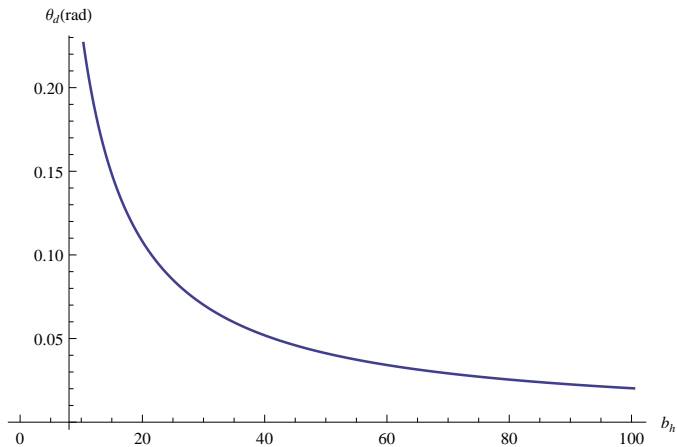}
\caption{The deflection angle as a function of the impact parameter in the classical GR solution.} 
\label{bvv1}
\end{figure}

\begin{figure}[t]
\centering
\subfigure[]{\includegraphics[scale=0.65]{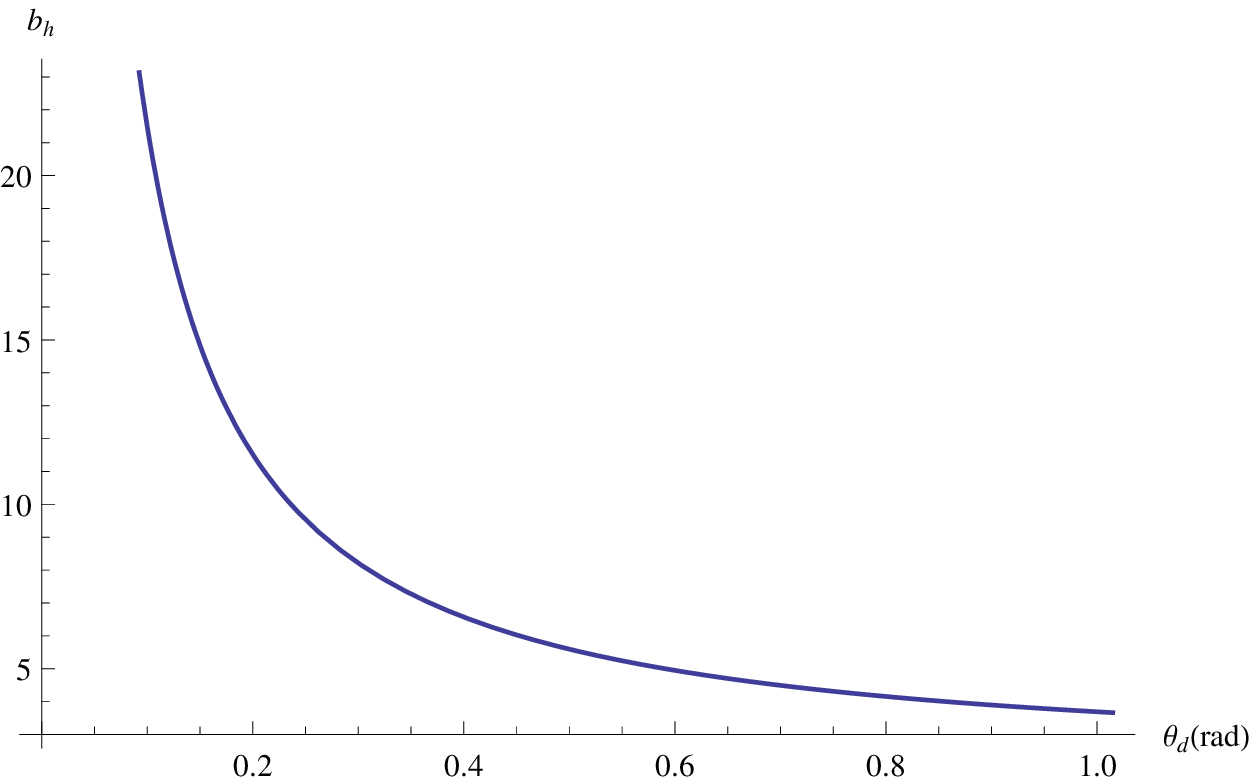}} \hspace{.5cm}
\subfigure[]{\includegraphics[scale=0.65]{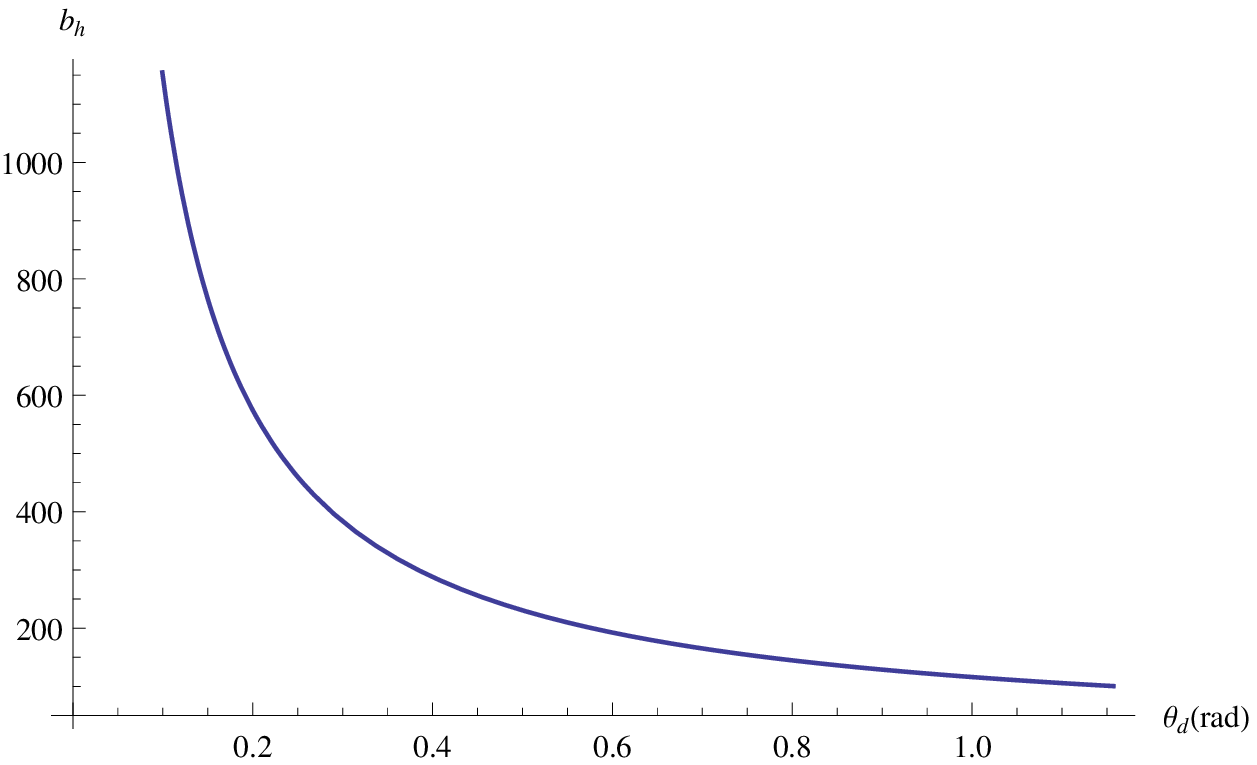}} \hspace{.5cm}
\subfigure[]{\includegraphics[scale=0.75]{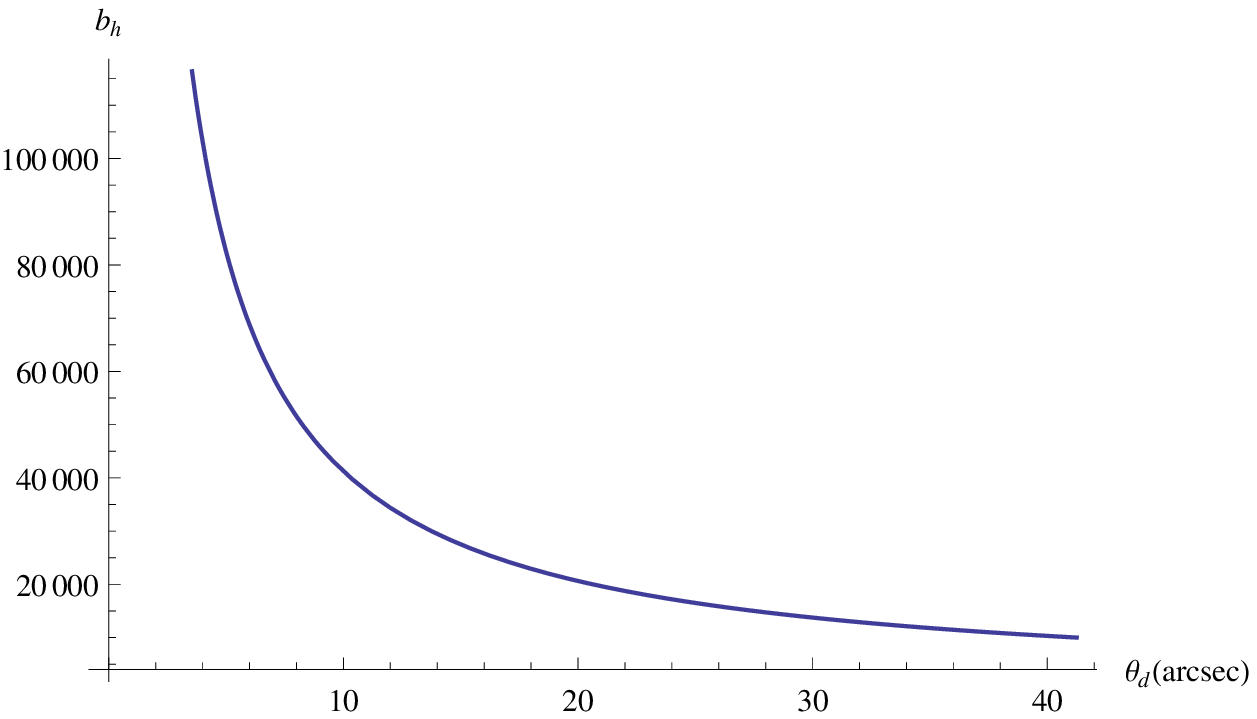}} \hspace{.5cm}
\caption{Fits of $b_h$ versus the deflection angle $\theta_d$ in the very near ( $b_h <\sim 20$), intermediate ($200 < b_h<  10^3$) and  distant ($20 \times 10^3< b_h < 10^5$) horizon regions.} 
\label{bvv}
\end{figure}

 We show in 
 Fig.\ref{bvv1} a plot of $\theta_d$ as a function of $x_0$, obtained by a numerical integration of the deflection formula Eq.(\ref{exactT}).
We also show in Fig.\ref{bvv} three fits of the GR expression for the deflection in the various regions using a functional expression of the form 
\beqa
b_h(\theta_d)&=& c_0 + \frac{c_1}{\theta_d} + \frac{c_2}{\theta_d^2} + c_3 \log \theta_d + c_4 \log \theta_d^2,
\label{func}
\eeqa   
with coefficients $c_i$ which depend on the three intervals in $\theta_d$ corresponding to the  
$( 3 < x_0 < 10^2)$, ($10^2 < x_0 < 10^4$) and  ($10^4 < x_0 < 10^6$) horizon regions. We have performed fits of the deflections to the same functional expression (\ref{func}), but using different numerical coefficients, which are in very good agreement with the direct numerical result obtaind from Eq.(\ref{erel}), and can be found in the appendix.

\subsection{The semiclassical relation at Born level}
\begin{figure}[t]
\centering
\subfigure[]{\includegraphics[scale=0.9]{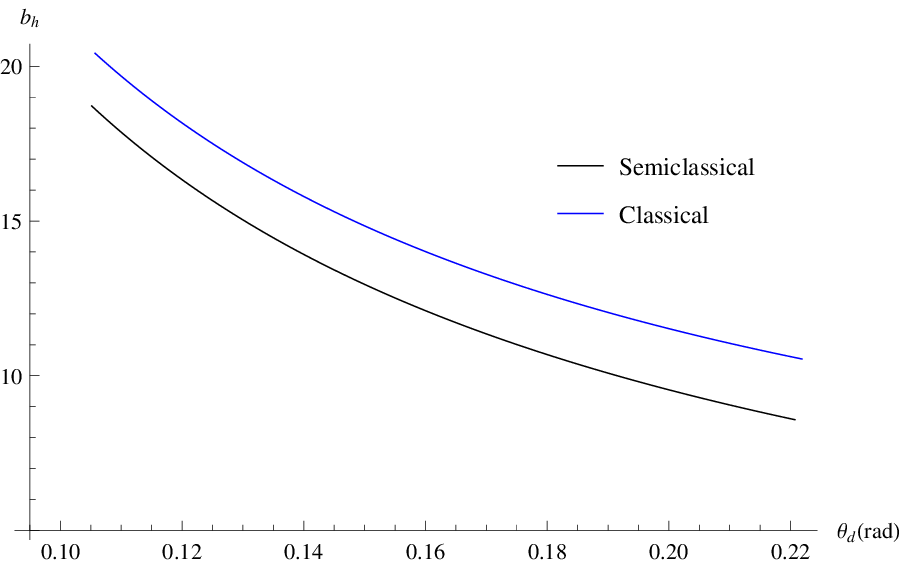}}
\subfigure[]{\includegraphics[scale=0.9]{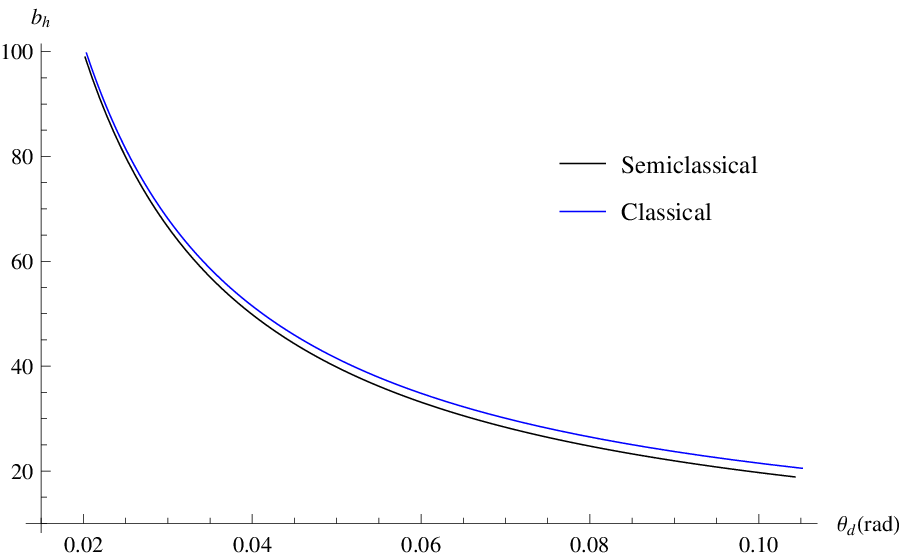}} 
\subfigure[]{\includegraphics[scale=0.9]{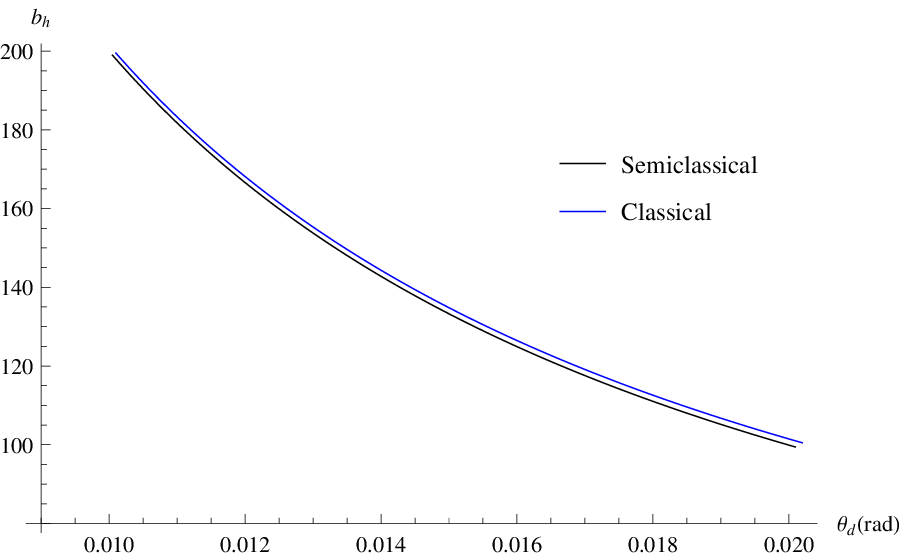}} \hspace{.5cm}
\caption{ Plots of the $b_h(\theta_d)$ for the GR and for the semiclassical result relation (Born level) for various impact parameter regions. Shown are the very near horizon region (top-left), the near horizon region (top-right) and the intermediate region (bottom). } 
\label{compare1}
\end{figure}

In order to compare the GR and the semiclassical result we turn to the semiclassical relation Eq.(\ref{semic}), with the cross section at the right hand side of this equation evaluated at Born level. The solution of Eq.(\ref{semic}) takes the form 
\beq
b_h^2({\theta_d})=b_h^2(\bar{\theta}) +2\int_{\theta_d}^{\bar{\theta}} d\theta' \sin\theta' \frac{d \tilde\sigma}{d\Omega'}, 
\label{intg}
\eeq
with $b_h^2(\bar{\theta})$ denoting the constant of integration. We have also denoted with $\tilde{\sigma}$ the unpolarized cross section stripped of the $(G M)^2$ prefactor, which has been removed in order to rewrite Eq.(\ref{semic}) 
in terms of $b_h$ rather than $b$.
The integration constant can be fixed by the condition that $b(\theta_d)$ approaches the classical GR solution as $b_h$ goes to infinity.  In this respect one can check by inspection that this condition is equivalent to the boundary condition 
\beq
\lim_{\theta_d \rightarrow \pi} b^2(\theta_d) = 0 \, , 
\label{bici}
\eeq
at zero impact parameter. This clearly shows that the requirement that the semiclassical equation reproduces the classical Einstein result at large impact parameters, selects a solution which is inadequate to describe the deflection close to the photon sphere. In fact, for a localized source surrounded by a horizon, Eq.(\ref{bici})
is not justified.  On the other hand, even for a distributed homogenous source, if we considered a beam moving towards the source with $b\sim 0$, the field felt by the  beam would be almost negligible, due to Gauss' law, and not strong enough to induce a photon to backscatter with an angle $\theta=\pi$, after a turn around the center of the source.\\  
There are few more comments which are in order when dealing with this relation between the classical and the quantum prediction for the deflection. It is clear that the notion of impact parameter is not part of the quantum description of a certain scattering process, but the appearance of $\theta_d$ as one of the extrema of the integration region in Eq.(\ref{intg}) carries a rather simple interpretation. In fact, in order to obtain a semiclassical picture of the angle of deflection we associate, for each given value of $\theta_d$, all the probabilities for a photon to scatter with $\theta^\prime >\theta$ up to the maximum value, $\theta^\prime=\pi$. Starting from this simple interpretation, the algorithm to be followed in order to link the quantum and the classical descriptions of the scattering process can be easily stated as follows.\\
1. We first identify the regions of impact parameter which we are interested in. The entire range of impact parameters, as we have already mentioned, can be split into several regions. The first region is the one which is {\em very near} to the photon sphere/horizon $( 3<x_0 < 20)$, followed by the {\em near horizon} region $( 20<x_0 < 100$), the {\em intermediate} region $( 100<x_0 < 10^4)$ and, finally, by the {\em far/distant} region $( 10^4<x_0 < 10^6)$. The latter describes the region which has been investigated in previous analysis  \cite{Berends:1975ah}. The sizes of these regions may differ, according to the observables that we need to investigate numerically, but this partition is basically preserved over the entire 
analysis.\\ 
2. We proceed by selecting two values for the closest distance between the source and the beam $x_0$, $(x_{0\,low}, x_{0 \,high})$,  corresponding to the interval of the impact parameter that we intend to investigate, using Eq.(\ref{xb}) to relate $b_h$ to $x_0$. \\
3. We then use the GR relation Eq.(\ref{erel}) to define the two classical angular values $\theta_{low}=\theta_d(x_{0\, high})$ 
and $\theta_{high}=\theta_d(x_{0\, low})$ within which we want to study the classical/quantum relation for the angular deflections. The differences between the two classical impact parameters in the selected angular region or, equivantly, 
 $(x_{0\,low}, x_{0 \,high})$  region, is then determined by the relation 
 \beq
 b_h(\theta_{low})-b_h(\theta_{high})= 2\int_{\theta_{low}}^{\theta_{high}} d\theta' \sin\theta' \frac{d \sigma}{d\Omega},
\eeq
which is then computed numerically. We illustrate the approach starting from Born level.
 At this level, from Eq.\ref{semic}) we obtain the differential equation
\beq
\frac{d b_0^2}{d\theta}= - 2\,  \left(G\,M\right)^2\, \cot^4\left(\frac{\theta}{2}\right)\, \sin\theta\, ,
\label{SquaredImpact}
\eeq
with $b_0$ denoting the value of $b$ computed at this order. The equation is separable and determines $b_0$ as a function of $\theta_d$, modulo an integration constant.  If we set this constant to zero we obtain the solution 
\beq
b_0^2(\theta_d)= 4\, G^2 M^2\, \left( \csc^2\left(\frac{\theta_d}{2}\right) + 4\, \log\sin\left(\frac{\theta_d}{2}\right)
  \label{bvt}                                                   - \sin^2\frac{\theta_d}{2}       \right)\, .
\eeq
already presented in \cite{Berends:1975ah}. In the small $\theta_d$ limit (i.e. for large angles of deflection) the solution above becomes
\beq
b_0 \sim G M\, \left(\frac{4}{\theta_d} +  \frac{\theta}{6} \left( 1 + 12\,\log \frac{\theta_d}{2} \right) \right )\, ,
\label{blocal}
\eeq
which allows us to identify the deflection angle $\theta_d$ as
\beq
\theta_d \sim \frac{4 \,G\, M}{b_0}\, ,
\label{impact}
\eeq
in agreement with the classical GR result. We will now proceed to investigate numerically the exact (brute force) solution of Eq.(\ref{semic}) at one-loop in the electroweak theory, and show that it agrees very accurately with Einstein's solution (\ref{erel}). The agreement is already very good at small values of the impact parameter ($b\sim 20 R_S$), and obviously extends out to infinite distance from the scattering center. As we have already mentioned in the Introduction, the numerical analysis shows that perturbation theory is significant not only in the description of the scatterings of photons which approach the scatterer at distances of the order of the solar radius or larger $(b\sim 0.5\times 10^6 R_S)$, but even at much smaller ones. Furthermore, by expressing the result in terms of $b_h$, the solution $b_h(\theta_d)$ shows not to depend on the size of the source, at least in leading order in $G M$. Obviously, this result would be modified by the quantum gravitational corrections, since these will induce a dependence on $GM$ which would not appear just as an overall factor in front of the one-loop cross section as in Eq.(\ref{csSM}), and the structure of the semiclassical equation will be much more involved.  \\
The closeness between the classical GR prediction and the semiclassical one is quite evident from Fig.\ref{compare1}, where we plot $b_h$ versus $\theta_d$ in different regions of the impact parameter $b_h$. It is quite clear that the two approaches predict similar deflections for $20 < b_h< 10^6$, with differences which start getting relevant only very close to the horizon (top-left panel) and hence concern situations of very strong lensing. 

\begin{figure}[t]
\centering
\includegraphics[scale=0.9]{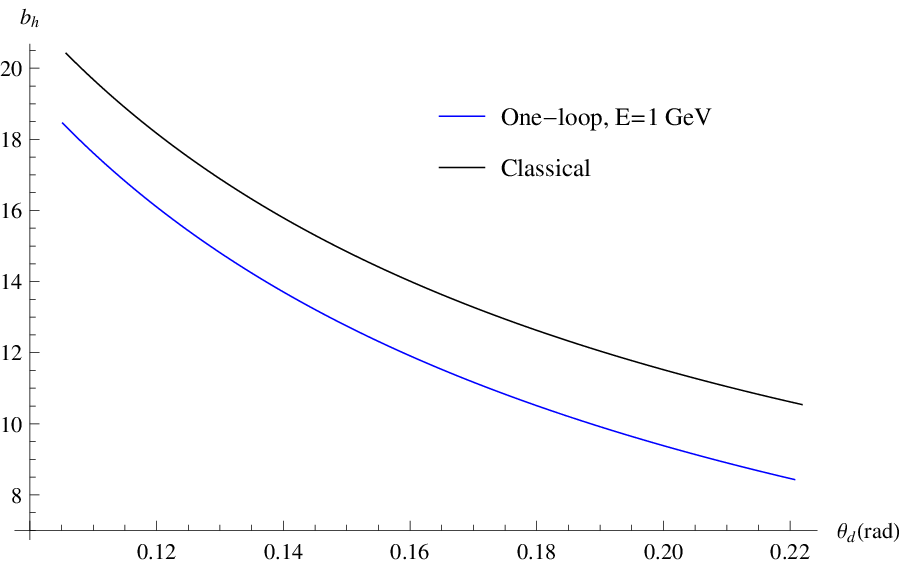}
\includegraphics[scale=0.9]{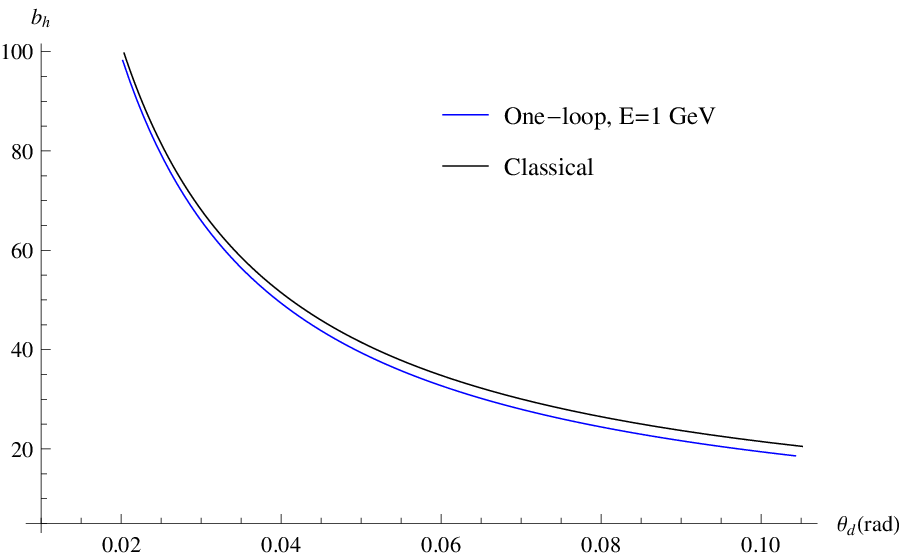}
\includegraphics[scale=0.9]{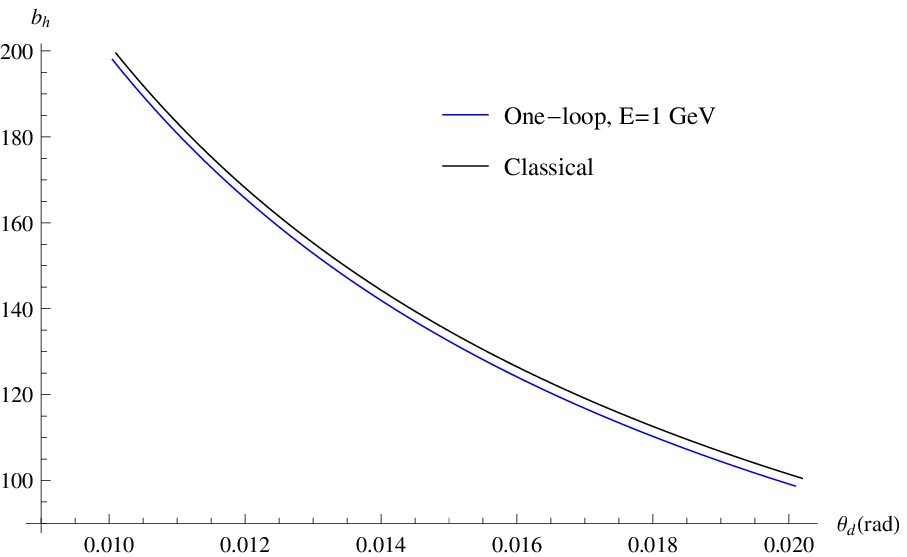}
\caption{Plots of the function $b_h(\theta_d)$ at one-loop in the SM, together with the classical GR results. Top left panel: the region  with $b_h < 20$. Top right panel: the region with $20< b_h < 100$. Bottom panel: the region with $b_h> 100$. }
\label{comploop}
\end{figure}

\begin{figure}[t]
\centering
\subfigure[]{\includegraphics[scale=1.2]{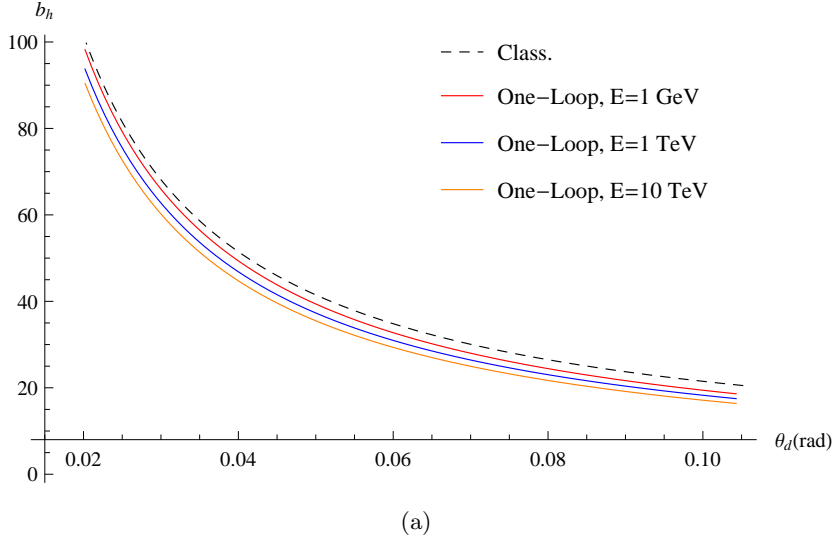}}
\caption{ Plots of the $b_h(\theta_d)$ for the GR and for the semiclassical result relation at one-loop level in the region 
$20 < b_h < 100$ from the high energy (1 GeV) to the very high energy region (10 TeV) of the gamma ray.} 
\label{compare2}
\end{figure}

\begin{figure}[t]
\centering
\includegraphics[scale=0.90]{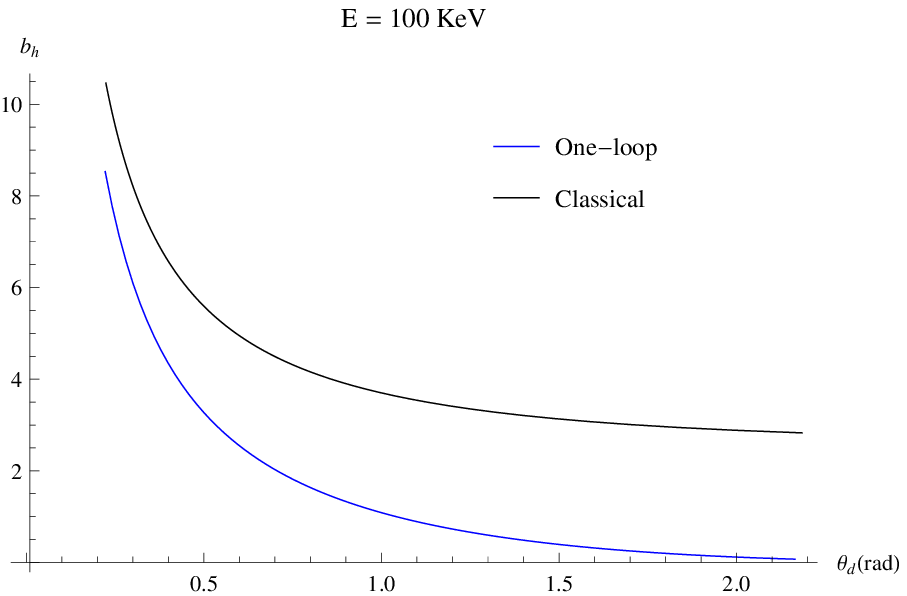}
\includegraphics[scale=0.90]{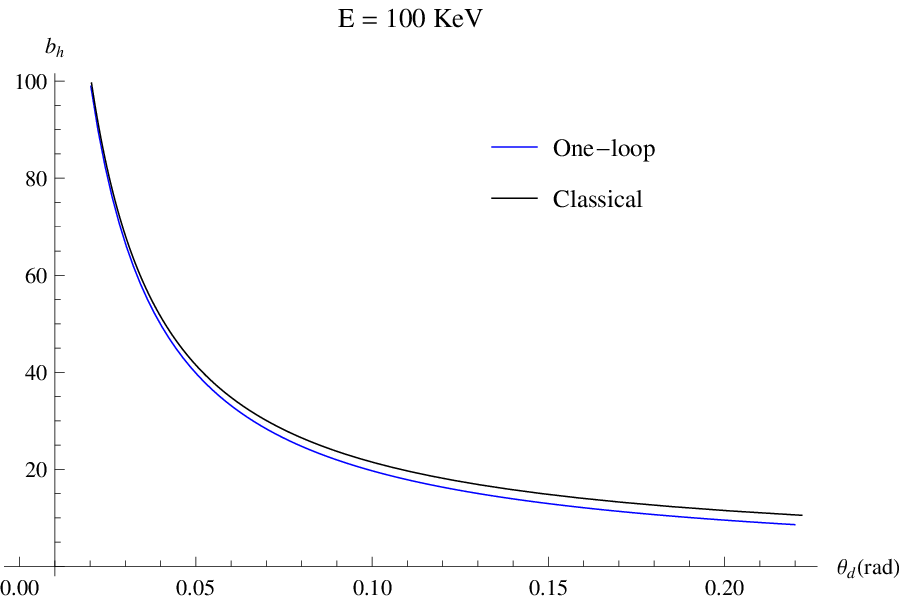}
\includegraphics[scale=0.9]{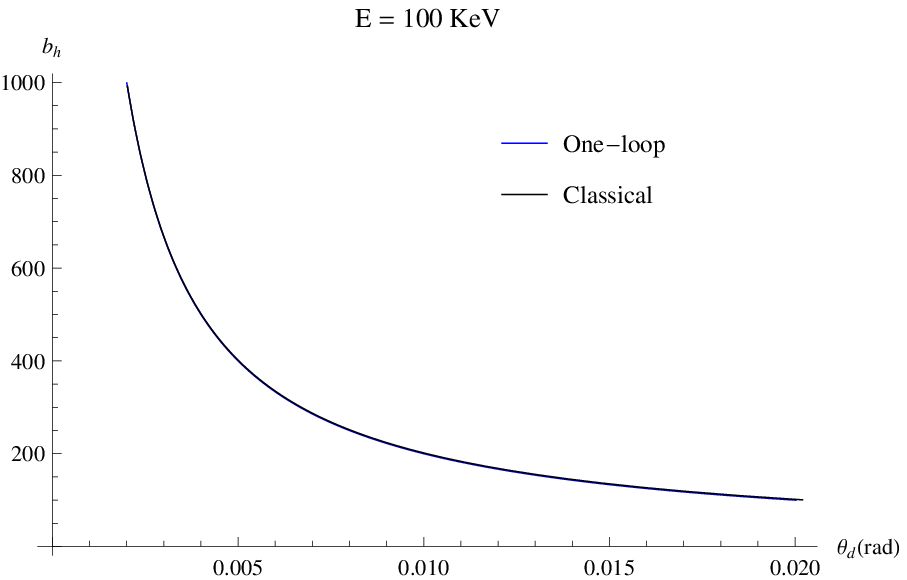}
\caption{Comparisons between the prediction for the relation $b_h(\theta_d)$ obtained from the GR formula versus the one-loop result with form factors expanded in the infrared region, with $t/m_e^2 \ll1$. }
\label{highE1}
\end{figure}

\begin{figure}[t]
\centering
\includegraphics[scale=0.90]{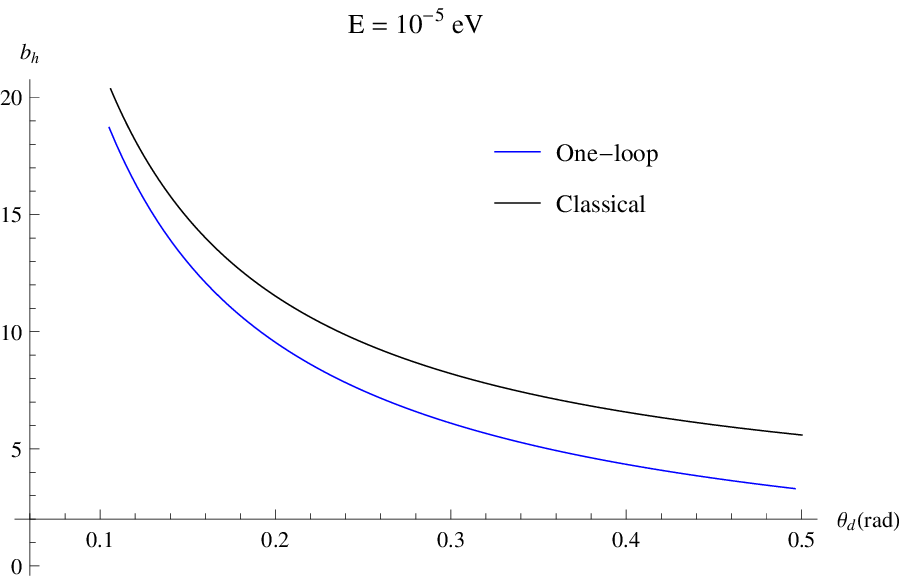}
\includegraphics[scale=0.90]{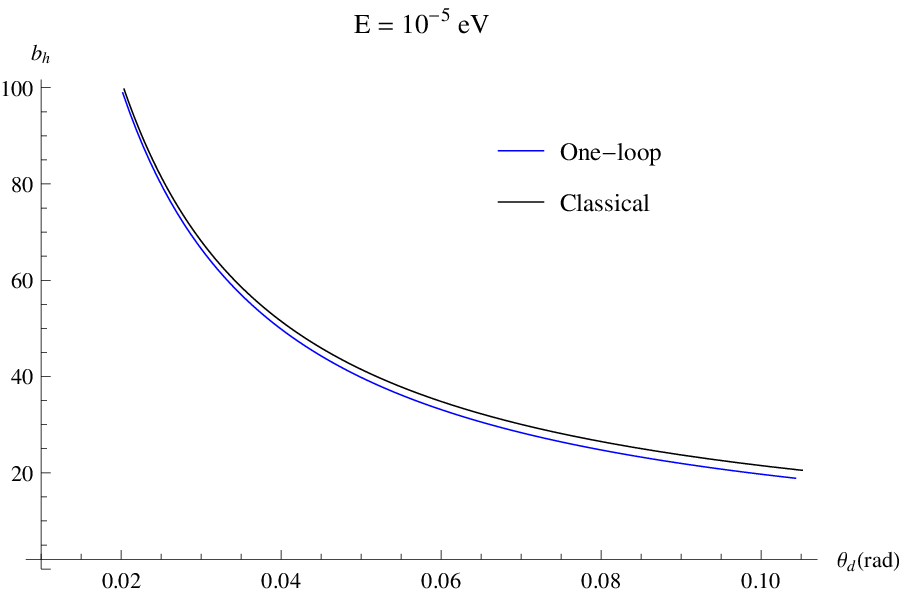}
\caption{Comparisons between the prediction for the relation $b_h(\theta_d)$ obtained from the GR formula versus the one-loop result with form factors expanded in the infrared region, with $t/m_e^2 \ll 1$ and CMB photons. }
\label{highE2}
\end{figure}

\subsection{The semiclassical relation at one-loop}
As we move to one-loop level, the numerical analysis requires particular care, due to the singular behaviour 
of the integrand of Eq.(\ref{intg}). This shows up especially at low energy, and motivates our choice to discuss separately 1) scatterings involving the incoming photon energy of high energy, of the order of 1 GeV and higher, from 2) those of lower energy, of the order of 1 MeV, down to the CMB case $\sim 10^{-5}$  eV. In the latter case we need to perform an expansion at small momentum transfers of the QED and weak form factors, which are analytic in the same limit, and perform the integration analytically. We show in Fig.\ref{comploop} three plots of the impact parameter $b_h$ against the angle of deflection at one-loop and we compare them with the classical GR prediction. As in previous plots, we split the impact parameter regions into the intervals $(10-20)b_h$, $(20-100) b_h$ and the region with $b_h > 100$, which illustrates the gradual overlap between the GR formula and the one-loop result as $b_h > 20$. Again, scatterings with impact parameters values below $20\, b_h$ should not be considered as correctly predicted from the semiclassical approach. 
The corrections induced by the radiative contributions to the angle of deflection are clearly rather small and are slightly more significant as we approach the horizon, towards $b_h\sim 20$, as shown by the tail behaviour of the top-right panel. 
Moving from 1 GeV to 10 TeV, see Fig.\ref{compare2}, also the relation between the angle of deflection and the impact parameter gets slightly modified. For instance, for an impact parameter of the order of $30\, b_h$
the one-loop and the GR result predict deflections which differ approximately by $10\%$ for a very energetic gamma in the TeV region. They remain confined, for the rest, at a few percent level at a lower energy, for incoming photons in the GeV range.

\subsection{Expansions at lower energy}%
As we have mentioned above, the study of photons characterized by an energy in the MeV region and below requires a special attention, due to severe numerical problems in the integration of Eq.(\ref{intg}). 
It is convenient, at this stage, to perform a $t/m^2$ expansion of the electroweak form factors 
in order to obtain an accurate solution of Eq.(\ref{semic}). Within this approximation, it is possible to obtain analytic solutions of the differential equation, determining the analytic expression of the function $b_h(\theta_d)$. Notice that, in this expansion, $m$ stands for the mass of any of the physical particles (fermions and gauge bosons) in the expressions of the loop corrections. For this reason, we require that the energy of the incoming beam is chosen in such a way that  $t/m_e^2 \ll1$, with $m_e$ being the electron mass. 
At leading order in $t/m_e^2$ the expressions of the form factor $\bar \Phi_3$, contributing to the unpolarized cross section, simplifies considerably as
\bea
\overline\Phi_{3\,QED} &=& - \sum_f N_c^f \frac{\alpha}{\pi}\, \frac{11\,Q_f^2}{90}\, \frac{E^2}{m_f^2}\, \sin^2\frac{\theta}{2} + O(t^2/m_e^4)\, , \nn \\
\overline\Phi_{3\,B} &=& - \frac{\alpha}{\pi}\, \frac{7}{5}\, \frac{E^2}{M_W^2}\, \sin^2\frac{\theta}{2}+ O(t^2/m_e^4) \, . 
\eea
Consequently, in the same limit the cross section becomes
\beq
\frac{d \sigma}{d \Omega}= G^2\, M^2\, \cot^4\left(\frac{\theta}{2}\right) \, \bigg[  1 - \frac{\alpha}{\pi}\, E^2\,  \bigg( \sum_{f} N_c^f\, \frac{11\,Q_f^2}{45\, m_f^2} 
+ \frac{14}{5\, M_W^2} \bigg)\, \sin^2\frac{\theta}{2} \bigg] + O(t^2/m_e^4).
\eeq
Denoting with $b_{h\, IR}(\theta_d)$ the impact parameter in the infrared limit and integrating Eq.(\ref{intg}),  we obtain for this function at leading order in $E/m_e$ the expression
\beqa
b_{h\, IR}(\theta_d)&=&\frac{1}{2} \left(\cos (\theta_d )+2 \csc ^2\left(\frac{\theta_d }{2}\right)+8 \log
   \left(\sin \left(\frac{\theta_d }{2}\right)\right)-1\right) \nonumber \\
&+&  \frac{\alpha}{\pi} E^2 \left( \sum_f \frac{11 N_c^f \,  Q_f^2}{720  {m_f}^2} + \frac{7 }{40 \pi  {M_W}^2} \right) \left(12 \cos (\theta_d )+\cos (2 \theta_d )+32 \log \left(\sin
   \left(\frac{\theta_d }{2}\right)\right)+11\right).\nonumber \\
\label{cmb}
\eeqa
We show in Fig.\ref{highE1} a comparison between the prediction for $b_{h\, IR}(\theta_d)$ and the GR formula. In this case we have expanded the form factors in the infrared region $b_h(\theta_d)$, with $t/m_e^2 \ll 1$, assuming an energy of the incoming beam of 100 KeV. Also in this case, as in the previous ones, the semiclassical and the classical result start to overlap from values of the impact parameter of the order of $20\, b_h$ (top-right panel). Clearly there are significant differences between the two predictions very close to the photon sphere, with  $10 < b_h < 20$ (top-left panel). On the other hand, the two results match completely as soon as we reach $100\, b_h$, showing that the radiative corrections become negligible already at such distances from the scattering center 
(Fig.\ref{highE1}, bottom panel). 

A similar behaviour is found also for the photons of the CMB. The cosmic background radiation pervades the universe and interacts gravitationally like any other form of radiation with the curved spacetime background. The typical wavelength at the peak of the distribution of the CMB is about $6 $ cm$^{-1}$, corresponding to a photon of $10^{-5}$ eV. Also in this case we can investigate the correspondence between the GR and the semiclassical prediction using an expansion of the semiclassical formula at $O(t/m_e^2)$. A direct numerical analysis shows that Eq.(\ref{cmb}) can be applied also to this case and predicts gravitational scattering and lensing of the photons at these wavelengths which is in agreement with the analysis of the previous sections. This is shown in Fig.\ref{highE2}, where we have investigated the relation between the classical GR deflection and the quantum prediction obtained from the SM. The two curves deviates for impact parameters below the $20\, b_h$ limit (left panel) but they basically overlap above $60\, b_h$. 
In the region between $20\, b_h$ and $60\, b_h$ the two curves present small differences (right panel), and are superimposed for larger values of the impact parameter. 
\subsection{Massless fermion limit and the conformal anomaly form factor} 
Before coming to our conclusions, we turn to a brief discussion of the contribution of the conformal anomaly to the deflection of a photon beam, working out the case of QED in the massless fermion limit. 
In this limit QED is conformally invariant and developes a conformal (trace) anomaly, which is present in those diagrams containing one or more insertions of the EMT, such as the $TAA$ vertex. As discussed in 
\cite{Giannotti:2008cv,Armillis:2009pq}, the anomaly is associated to the apperance of a massless pole in a specific form factor, which in our notations is denoted by $\overline\Phi_{1,F}$ (see (\ref{IncludeAmp2})), the anomaly form factor. This is characterized by a massless pole and by a typical sum rule 
\cite{Giannotti:2008cv} which has been recently studied also in the context of supersymmetric theories \cite{Coriano:2014gja}. The massless exchange is viewed, at the level of the effective action, as describing a scalar composite interpolating field, triggered by the gravitational interaction. 
The goal of this section is to investigate at a phenomenological level the possible implications of the anomaly contribution to the deflection of light. We are going to show that the effect of this interaction is to induce a helicity flip of the incoming photon beam. We will see that the contribution of the anomaly to the $(+ -)$ cross section appears at $O(\alpha^2)$, and differs from the Born level $(++)$ cross section just by an overall factor. For this purpose we recall that the $(++)$ cross section coincides with the unpolarized cross section, as shown in (\ref{firstorder}), and hence the anomaly contribution can be viewed as a renormalization of the unpolarized result. 

To better clarify these points,  for simplicity we consider the case of a single massless fermion running in the loop and coupled to the incoming and outgoing photons. A direct computation shows that in this limit the form factors are given by 
\bea
\overline\Phi_{1,F}(0,0,0,0)=-\frac{2\,\alpha\, Q_f^2}{9\,\p \,t} \,, \qquad \overline\Phi_{2,F}(0,0,0,0)=-\frac{\alpha\, Q_f^2}{36\,\p\, t} \,.
\eea
The expressions above, inserted into Eq.(\ref{polplusminus}) give 
\bea
\frac{d \sigma}{d \Omega}^{+-}&=&\frac{\alpha ^2 (G M)^2 {Q_f}^4}{36 \pi ^2} \cot ^4\left(\frac{\theta}{2}\right) = A^2 \frac{d \sigma}{d \Omega}_0
\label{2loop}
\eeqa
with
\beq
A\equiv \alpha\frac{ Q_f^2}{6 \pi}= 3.87 \times 10^{-4} Q_f^2,
\eeq
and is proportional to the unpolarized tree-level cross section given in Eq.(\ref{leading}).\\
To quantify the size of the contribution of the $(+-)$ amplitude - and hence of the conformal anomaly - to the unpolarized cross section, we start by recalling that this cross section is of the form given by Eq.(\ref{firstorder}) at $O(\alpha)$, while its complete expression at $O(\alpha^2)$ remains unknown. 
We denote with $\left.\frac{d \sigma}{d \Omega}^{++}\right|_2$ the unknown contribution to the (++) sector of the unpolarized $\frac{d \sigma}{d \Omega}$ at $O(\alpha^2)$, so that this can be re-expressed in the form 
\beq
\frac{d \sigma}{d \Omega}=\left(\frac{d \sigma}{d \Omega}^{++} +\left.\frac{d \sigma}{d \Omega}^{++}\right|_2 + \frac{d \sigma}{d \Omega}^{+-}\right),
\label{mod}
\eeq
In the equation above, we have denoted with $(d \sigma/{d \Omega})^{++}$ the (known) $O(\alpha)$ result of the previous sections, while 
 $(d \sigma/{d \Omega})^{+-}$ is given by Eq.(\ref{2loop}). To investigate the contribution to the lensing of photons of (\ref{mod}) we  rewrite the semiclassical relation Eq.(\ref{semic}) at $O(\alpha^2)$ split in the following form 
\bea
b_{++}^2(\theta_d) &=& 2\int_{\theta_d}^{\pi} d\theta' \sin\theta' \frac{d \sigma}{d\Omega}^{++}\nonumber, \\
b_{++ 2}^2 (\theta_d)&=& 2\int_{\theta_d}^{\pi} d\theta' \sin\theta'\left. \frac{d \sigma}{d\Omega}^{++}\right|_2\nonumber ,\\
b_{+-}^2 (\theta_d)&=& 2\int_{\theta_d}^{\pi} d\theta' \sin\theta' \frac{d \sigma}{d\Omega}^{+-}.
\label{intg1}
\eeqa
with 
\beq
b^2=b_{++}^2 + b_{++ 2}^2 + b_{+-}^2.
\eeq
Neglecting for the moment the contribution to the deflection coming from $b_{++ 2}^2$, we focus on the related $b_{+-}^2$ value, which is proportional to the tree-level contribution, as clear from Eq.(\ref{2loop}). Therefore, by combining this result with the tree-level one, given by the same equation, the change in the formula of deflection amounts to a renormalization by a factor $(1+ A^2)$ 

\beq
b_{h,0}^2(\theta_d)\to (1+A^2)  \left( \csc^2\left(\frac{\theta_d}{2}\right) + 4\, \log\sin\left(\frac{\theta_d}{2}\right)
  \label{bvt1}                                                   - \sin^2\frac{\theta_d}{2}       \right).\, 
\eeq
The asymptotic deflection, using the fact that $A \ll1$, takes the new form 
\beq
b \sim G M (1+\frac{A^2}{2})\, \left(\frac{4}{\theta_d} +  \frac{\theta_d}{6} \left( 1 + 12\,\log \frac{\theta_d}{2} \right) \right )\, 
\label{blocal1}.
\eeq
It is clear that the classical Einstein relation Eq. (\ref{impact}) is given by
\beq
\theta_d= \frac{4 G M}{b}(1 + \frac{A^2}{2}) +...
\eeq
where the ellipses refer to the additional (unknown) $O(\alpha)$ contributions. These are terms obtained from the interference between the tree-level and the one-loop Feynman graphs, together with the remaining unknown $O(\alpha^2)$ corrections coming from 
$(d \sigma/{d\Omega})^{++}|_2$. Notice that $A\sim 10^{-4}$ for a single fermion, and therefore the anomaly form factor modifies the angular deflection by few parts per thousands, compared to the Einstein prediction. Notice also that this result remains valid both for weak and for very strong lensings, as clear from (\ref{bvt1}).

\section{Conclusions}
We have presented the explicit expression of the radiative corrections to the cross section describing the scattering of a high energy photon over a spherically symmetric gravitational background, in the SM. We have shown that these corrections become quite sizeable in the very high energy region.
We have also performed a comparative analysis of the classical and of the quantum approaches to the lensing of a photon. This comparison has been performed by equating the classical cross section - defined in terms of the impact parameter of the photon - and the ordinary perturbative cross section computed using the Lagrangian of the SM. This semiclassical approach allows, in general, to derive a relation between the impact parameter $b$ and the deflection angle $\theta_d$, by solving a differential equation.\\
 Our numerical analysis shows that the weak field expansion for gravity is extremely effective in the description of the lensing of a photon from the far distant down to the near horizon region, with no need to include extra corrections proportional to the curvature of the background. The remarkable agreement between the classical (GR) and the semiclassical approaches holds down to distances of about 20 horizon units ($20\, b_h$), in conditions of strong lensing, proving that pertubation theory is remarkably effective in the description of strong lensing. Our analysis extends considerably previous studies of the quantum corrections to the deflection of photons, which have been limited, in the past, only to regions of very large impact parameters ($b_h\sim 10^6$). \\
  We have then studied the helicity-flip cross section, which is radiatively induced, and plays an important role in the determination of a change in the polarization of a photon beam in background gravity. In the case of the CMB these effects are extremely small for a black hole of the order 1.4 solar masses, but grow significantly with the mass of the source and with the energy of the photons. However, the helicity conserving amplitude remains the dominant part of the unpolarized cross section.\\
   We have also investigated the same de-polarizing amplitude and related cross section in the massless fermion limit, for QED. This theory is affected by the conformal anomaly. Both for strong and weak deflections, the contribution of the anomaly form factor is at the level of $\sim 3\times 10^{-4}$ in all the angular regions, for each massless fermions included in the virtual corrections. 
It is proportional to the tree-level result, which is energy-independent. \\
Our analysis can be extended in various directions, one of the most significant being the possible implications of our results in the neutrino sector. We hope to return in the near future with a discussion of this and of other related issues not addressed in this work.

\centerline{\bf Acknowledgements} 

We thank Paolo Amore and Pietro Colangelo for discussions.

\section{Appendix. Numerical fits of the GR deflection}
\appendix
The coefficients corresponding to the fit in Eq.(\ref{func}) are given by
\beqa
b^1_h(\theta_d)&=&\frac{0.000043695}{\theta_d ^2}+\frac{1.99202}{\theta_d }+0.0454744 \log ^2(\theta_d )+0.16025
   \log (\theta_d )+1.69987 \,, \nn \\
b^2_h(\theta_d)&=& \frac{9.93100299967969\times 10^{-11}}{\theta_d^2}+\frac{2.}{\theta_d
   }+0.000740459 \log ^2(\theta_d )+0.00896133 \log (\theta_d )+1.50095  \,, \nn \\
b^3_h(\theta_d)&=& \frac{2.006051361087774 \times 10^{-11}}{
\theta_d^2}+\frac{1.99998}{\theta_d }+0.28887 \log ^2(\theta_d )+5.34226 \log (\theta_d )+26.3238 \,. \nn\\
\eeqa
The regions covered by the fits correspond to the intervals (expressed in radians) 
$(0.0201967, 1.01488)$, $(0.000200019, 0.0201967)$ and $(2.00001\times10^{-6}, 0.000200019)$ 
respectively.




\end{document}